\documentclass[11pt,letterpaper]{article}
\pdfoutput=1
\usepackage{jheppub}
\usepackage{natbib}

\usepackage{tikz}
\usetikzlibrary{calc}

\usepackage{hyperref}
\usepackage{amsmath}
\usepackage{amsfonts}
\usepackage{graphicx}
\usepackage{caption}
\usepackage{placeins}
\usepackage{tqft}
\usepackage{comment}
\usepackage{array}

\DeclareRobustCommand {\D}{\mathbf{D}}
\DeclareRobustCommand{\C}{\mathcal{C}}
\DeclareRobustCommand{\H}{\mathbb{H}}

\DeclareRobustCommand{\Aut}{\mathrm{Aut}}
\DeclareRobustCommand{\h}{\mathrm{H}}

\title{Hyperbolic Geometry and Amplituhedra in 1+2 dimensions}

\author[a,c]{G.~Salvatori}
\author[b,c]{S.~L.~Cacciatori,}

\affiliation[a]{Dipartimento di Fisica, Universit\`a degli Studi di Milano,
Via Celoria 16, IT-20133 Milano, Italy}
\affiliation[b]{Department of Science and High Technology, \\
Universit\`a dell'  Insubria, \\
Via Valleggio 11, I-22100 Como, Italy.}
\affiliation[c]{INFN, Sezione di Milano, \\
Via Celoria 16, I-20133 Milano, Italy.}

\emailAdd{giulio.salvatori@unimi.it}
\emailAdd{sergio.cacciatori@uninsubria.it}

\abstract{Recently, the existence of an Amplituhedron for tree level amplitudes in the bi-adjoint scalar field theory has been proved by Arkhani-Hamed et al. We argue that hyperbolic geometry constitutes a natural framework to address the study of 
positive geometries in moduli spaces of Riemann surfaces, and thus to try to extend this achievement beyond tree level. In this paper we begin an exploration of these ideas starting from the simplest example of hyperbolic geometry, 
the hyperbolic plane. The hyperboloid model naturally guides us to re-discover the moduli space Associahedron, and a new version of its kinematical avatar. As a by-product we obtain a solution to the scattering equations which can be interpreted as a 
special case of the two well known solutions in terms of spinor-helicity formalism. The construction is done in $1+2$ dimensions and this makes harder to understand how to extract the amplitude from the dlog of the space time Associahedron.
Nevertheless, we continue the investigation accommodating a loop momentum in the picture. By doing this we are led to another polytope called Halohedron, which was already known to mathematicians. We argue that the Halohedron fulfils many criteria that 
make it plausible to be understood as a 1-loop Amplituhedron for the cubic theory. Furthermore, the hyperboloid model again allows to understand that a kinematical version of the Halohedron exists and is related to the one living in moduli space by a simple generalisation of the tree level map.}

\begin{document}
\maketitle
\flushbottom

\section{Introduction}

In recent years we experienced a tremendous progress in the computation of scattering amplitudes for massless theories. In a virtuous loop these computations allowed to gain new conceptual tools that pushed the boundary of what is computable even further. 
The Amplituhedron \cite{Arkani-Hamed:2013jha,ArkaniHamed:2012nw} is a beautiful instance of these ideas, it is a generalised polyhedron that encodes in its \emph{positive geometry} the local and unitary properties of the scattering amplitudes in $\mathcal{N}=4$ planar Super Yang-Mills, which can now be 
computed as the ``Volume'' of this object. 

Whilst being a satisfying result in its own, the discovery of the Amplituhedron is by no means sufficient to replace - or better translate - everything we knew on quantum field theories with an equivalent dictionary in the world of positive geometries. Indeed, till 
very recently \cite{Arkani-Hamed:2017mur,delaCruz:2017zqr}, there was no other instance of a quantum field theory whose amplitudes could be computed in the same amplituhedral fashion as $\mathcal{N}=4$ SYM. However, we now know that this is possible for a larger class of theories, such as 
the \emph{bi-adjoint scalar} field theory, at least at tree level. The relevant Amplituhedron for this theory is called Associahedron, whose discovery in mathematics dates back to the 60's, and was already used in stringy computations of amplitudes. The fact that 
positive geometries pop out in theories as different as a scalar cubic theory and a maximally super symmetric one is an undeniable evidence that the Amplituhedron is not a ``quirk'' of a very peculiar theory. 

The Associahedron story is deeply connected with the Cachazo-He-Yuan formula \cite{Cachazo:2013gna,Cachazo:2013hca,Cachazo:2013iea,Cachazo:2016ror} that represents amplitudes as contour integrals on the moduli space of Riemann surfaces supported on the locus of the so called ``scattering equations''. The core of this 
connection is the property of the moduli space of factorising as an amplitude should, the existence of a differential form with logarithmic singularities and factorised residues and finally the fact that the scattering equations map singular kinematical regions 
in their correspective boundaries of the moduli space.

How these main ingredients can be put together to define an Amplituhedron for a theory was explained in \cite{Arkani-Hamed:2017mur}, where it was first defined in rigorous terms what a \emph{Positive Geometry} is. As the name suggests, the heart of a positive geometry has to 
do with real numbers, and indeed positive geometries are defined in a suitable real section of a complex manifold as the domain where certain functions are positive. Saturating this positiveness conditions we get the boundary of the positive geometry which 
have to be positive geometries themselves. In this sense, there is a plethora of positive geometries that ranges from objects as simple as a triangle up to the original Amplituhedron. The construction of the canonical dlog form - which is actually itself 
a definitional   property of the geometry - can be done in many ways, among which by push-forwarding a previously known dlog through an appropriate morphism. The CHY prescription can be described exactly as a pushforward from the Associahedron 
living in the moduli  space to a kinematical Associahedron. The fact that a positive geometry can be found directly in kinematical space is another remarkable feat achieved in \cite{Arkani-Hamed:2017mur}.

\begin{figure}[!htbp]
\begin{center}
\begin{tikzpicture}[tqft/flow=east, tqft/boundary style={fill=white,draw=black,opacity=0.5}]
\node[tqft/reverse pair of pants,draw] (a) {};
\begin{scope}[tqft/flow=east, tqft/boundary style={fill=white,draw=white,opacity=0}]
\node[tqft/cylinder, draw, opacity=0, anchor=incoming boundary 1](b) at (a.outgoing boundary 1){};
\end{scope}
\node[tqft/pair of pants,draw,anchor=incoming boundary 1](c) at (b.outgoing boundary 1){};
\path (b.outgoing boundary 1) ++(2.5,0) node[font=\Huge] {\(=\)};
\path (b.outgoing boundary 1) ++(5,0) node[tqft/reverse pair of pants,draw,anchor=outgoing boundary 1] (e) {};
\node[tqft/pair of pants,draw,anchor=incoming boundary 1] (f) at (e.outgoing boundary 1) {};
\end{tikzpicture}
\caption{A pair of pants decomposition of a sphere with four punctures.}
\label{fig:pairofpants}
\end{center}
\end{figure}
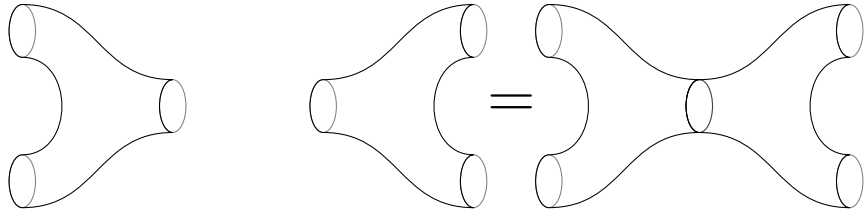

It is very natural that to pursue these ideas further in the computations at loop level, and deeper in the understanding of the role of positiveness in physics, we have to gain a better understanding on the real structures that can be found in moduli spaces at 
higher genera, where CHY formulae have already been studied. A natural starting point to do so is the hyperbolic approach to the moduli problem, which consists in building the most general hyperbolic Riemann surface by gluing elementary building blocks - the 
pair of pants - along their geodesic boundary, see figure \ref{fig:pairofpants}. A gluing is specified by the length of the geodesic boundary as well as the relative angle of the pants, this pair of real numbers form the Fenchel-Nielsen coordinates. A natural definition of a real section would be to take all the angles to be zero, and allow only the lengths to vary.

\FloatBarrier

Inequalities are ubiquitous in the hyperbolic approach and moduli arising as geodesic lengths usually have to satisfy interesting constraints, reminiscent of those that have been seen in the world of positive geometries. For example, take Rengel's famous 
inequality that bounds the width $W$ and height $H$ of an hyperbolic quadrilateral $Q$ of area $A$ to its modulus,
\begin{align*}
\frac{H^2}{A} \le \mathrm{Mod}(Q) \le \frac{A}{W^2}.
\end{align*}
Loosely speaking this inequality states that if we want to stretch an hyperbolic quadrilateral in one direction we have to squeeze it in the other. 
\begin{figure}[!htbp]
\begin{center}
\begin{tikzpicture}[>=latex]
\node at (-1,-2-0.4) {$\pmb {1}$};
\node at (1,-2-0.4) {$\pmb {2}$};
\node at (1,2.4) {$\pmb {3}$};
\node at (-1,2.4) {$\pmb {4}$};
\shade [left color=blue!30!white, right color=blue!10!white] (-1,-2) .. controls (-0.7,-1.7) and (0.7,-1.7) .. (1,-2) .. controls (0.5,-1.6) and (0.5,1.6) ..(1,2) .. controls (0.7,1.7) and (-0.7,1.7) .. (-1,2) .. controls (-0.5,1.6) and (-0.5,-1.6) .. (-1,-2) -- cycle;
\draw [blue!50!white, ultra thick] (-1,-2) .. controls (-0.7,-1.7) and (0.7,-1.7) .. (1,-2) .. controls (0.5,-1.6) and (0.5,1.6) ..(1,2) .. controls (0.7,1.7) and (-0.7,1.7) .. (-1,2) .. controls (-0.5,1.6) and (-0.5,-1.6) .. (-1,-2) -- cycle;
\begin{scope}[xshift = 5cm]
\node at (-2-0.4,-1) {$\pmb {1}$};
\node at (2+0.4,-1) {$\pmb {2}$};
\node at (2+0.4,1) {$\pmb {3}$};
\node at (-2-0.4,1) {$\pmb {4}$};
\shade [left color=blue!30!white, right color=blue!10!white] (-2,-1) .. controls (-1.5,-0.7) and (1.5,-0.7) .. (2,-1) .. controls (1.75,-0.6) and (1.75,0.6) ..(2,1) .. controls (1.5,0.7) and (-1.5,0.7) .. (-2,1) .. controls (-1.75,0.6) and (-1.75,-0.6) .. (-2,-1) -- cycle;
\draw [blue!50!white, ultra thick] (-2,-1) .. controls (-1.5,-0.7) and (1.5,-0.7) .. (2,-1) .. controls (1.75,-0.6) and (1.75,0.6) ..(2,1) .. controls (1.5,0.7) and (-1.5,0.7) .. (-2,1) .. controls (-1.75,0.6) and (-1.75,-0.6) .. (-2,-1) -- cycle;
\end{scope}
\end{tikzpicture}
\caption{Stretching of an hyperbolic quadrilateral. All the edges are geodetics on a hyperbolic Riemann surface.}
\label{fig:diagonalicross}
\end{center}
\end{figure}
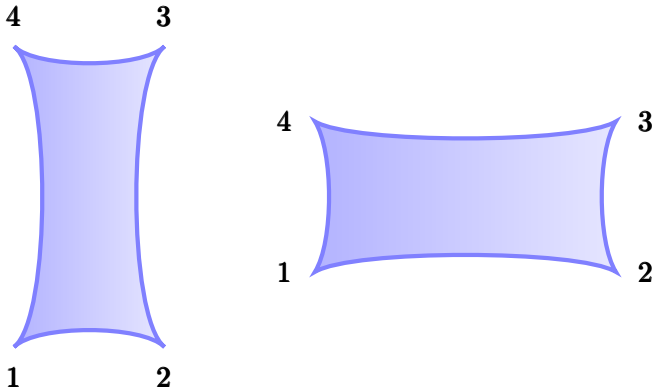
\FloatBarrier

\noindent This should be compared with the fixing $k_{13} \doteq \mathrm{const}$ used in defining the 4-point kinematical Associahedron, that forces an increase in $k_{12}$ to be compensated by a decrease of $k_{14}$.

In this paper we start to explore the relations with hyperbolic geometry and positive geometries relevant for scattering amplitudes. 

We begin with the simplest case of the hyperbolic plane which has a trivial connection with $1+2$ dimensional space time: the hyperboloid model. Already this well known fact has interesting consequences. 
On the one hand, the hyperbolic approach makes manifest that the moduli space for the hyperbolic plane with marked points at infinity is an Associahedron - indeed it is exactly the one considered by Arkhani-Hamed et al. On the other hand, the hyperboloid model allows to recognise this Associahedron directly in space time and provides a map from kinematics to the moduli space which can be understood as being a particular solution of the scattering equations. Moreover from this point of view we are immediately forced to think of the null momenta of the scattering particles as \emph{homogeneous coordinates} for the moduli space. This is interesting especially in light of the recent proposal of \cite{Arkani-Hamed:2017mur} that ``projectivity'' of the mandelstam variable is a crucial ingredient to pin down the planar scattering form and thus the amplitude. However, in our approach this fact seems to be a non trivial obstacle to extract the amplitude from the canonical dlog of the kinematical Associahedron, since the latter cannot be realised as a compact polytope in $\mathbb{R}^N$.

Probably the most interesting result described in this paper is the fact that the tree level Associahedron has a natural generalisation at 1-loop level: we simply have to cut a circle away from the hyperbolic plane and by taking the complex double we get a torus. Of course, in the same way as the tree Associahedron is not the entire moduli space $\mathcal{M}_{n}(\mathbb{R})$ where the scattering equations live, the surfaces we obtain cutting away a circle and doubling are not the most tori with $n$ markings. Instead the moduli obtained this way span a convex polytope called \emph{Halohedron}, well known to mathematicians \cite{devadoss}.
As we will see the Haloehdron geometry has a physical interpretation as it encodes the factorisations and cut properties of a $1$-loop integrand of the bi-adjoint scalar theory, and therefore it is a natural candidate for being the $1$-loop Amplituhedron. In addition, the hyperboloid construction used at tree level can readily be extended to accomodate a loop momenta in the kinematical data, so that it allows to identify an Halohedron directly in kinematical space and provides a map between the two Halohedra.
Obviously, the unsolved tree level problems plague this construction as well, therefore an explicit computation of the 1-loop integrand is not at hand as well as a conclusive proof of these statements.

This paper is organised as follows. In section \ref{sec:review}, for the reader's convenience, we provide a quick review of the geometry of the hyperbolic plane and some elementary facts on the hyperbolic approach to Riemann surfaces. In section \ref{sec:tree}, we study the tree level Associahedron, its kinematical counterpart and the map $\phi$ that links them. In section \ref{sec:loop} we explain how these ideas can be extended at 1-loop level, and we study the connections of the Halohedron with the $1$-loop planar integrand. Finally, in section \ref{sec:conclusions} we elaborate on some problems encountered along the way and suggest future directions of research.


\section{Review of hyperbolic geometry}
\label{sec:review}
Here we review some elementary facts on plane hyperbolic geometry, Riemann surfaces and moduli. Further details can be found in \cite{abikoff,hubbard}.
\subsection{Hyperbolic Plane}

The simplest and most fundamental hyperbolic surface is the hyperbolic plane, which can be thought of in many ways.
The first model of the hyperbolic plane is the \emph{Poincar\'e disk}, which is defined as the open disk $\mathbf{D} \subset \mathbb{C}$  of radius 1 centered at the origin. The bi-holomorphic maps $\phi : \mathbf{D} \to \mathbf{D}$ form its automorphism group $\mathrm{Aut}_\mathbf{D}$. Any element $\psi \in \mathrm{Aut}_\mathbf{D}$ is 
of the form 
\begin{align*}
\psi : z \mapsto \phi(z) = \frac{a z + b}{\bar{b} z + \bar{a}},
\end{align*}
where $a, b \in \mathbb{C}$ and $ |a|^2 > |b|^2$. In the conformal class of $\D$ there is a privileged metric, given by the infinitesimal distance
\begin{align*}
\rho_\D = \frac{2 |dz|}{1-|z|^2},
\end{align*}
which is left invariant by all the elements of $\mathrm{Aut}_{\D}$.
All Riemannian metrics of $\D$ compatible with this requirement are obtained by a global scalings of $\rho_\D$\footnote{The factor $2$ in the numerator of $\rho_\D$ is chosen to normalize the curvature of $\D$ to $-1$.}. In terms of this metric we can define 
objects invariant under automorphism, such as lengths, areas, geodesics, circles and so on. The Poincar\'e disk equipped with $\rho_\mathbf{D}$ is a Riemannian manifold called \emph{Hyperbolic plane}. We can characterize the elements of $\Aut_\D$, in terms of their fixed points and in terms of the metric and we divide them in parabolic, elliptic, and hyperbolic transformations:  
For each $\psi \in \Aut_\D$ define $m(\psi):= \inf_{z \in \D} \mathrm{d}(z,\psi(z))$, where $d$ is the geodesic distance. Then we have the following table.
\begin{center}
    \begin{tabular}{ | l | l | l | p{5cm} |}
     \hline
    Parabolic & One fixed point in $\partial{\D}$ & $m(\psi) = 0$, Infimum not obtained in $\D$ \\ \hline
    Elliptic & One fixed point in $\D$ & $m(\psi) = 0$, Infimum obtained in $\D$ \\ \hline
    Hyperbolic & Two fixed points in $\partial{\D}$ & $m(\psi) > 0$ \\ \hline
    \end{tabular}
\end{center}

There are many other models of the Hyperbolic plane, one with an immediate connection with the scattering amplitudes is the upper half plane model. The upper half plane is the set $\mathbb{H} \subset \mathbb{C}$ with $\mathrm{Im}(z)>0$. Since it is simply 
connected, it must be bi-holomorphic to $\D$, and, indeed, there is a very well known map,
\begin{align*}
\mathcal{C}:\ &\D \to \mathbb{H} \\
&z \to \sigma = \frac{i(z+1)}{1-z},
\end{align*}
called the (inverse) Cayley transform. $\C$ is a Moebius map which sends the boundary circle of $\D$ to the real axis. The automorphism group $\Aut_\mathbb{H}$ is given by
\begin{align*}
\phi : z \mapsto \phi(z) = \frac{a z + b}{c z + d},
\end{align*}
with $a,b,c,d \in \mathbb{R}$ and with $a c - b d > 0$. We can identify $\Aut_\mathbb{H} \approx \mathrm{PSL(2,\mathbb{R})} = \mathrm{SL(2,\mathbb{R})}/\{\pm \mathrm{I}\}$.
The reader familiar with the scattering equations literature will notice that the real section of the moduli space used in \cite{Arkani-Hamed:2017mur} to define a positive geometry is very close to $\H$. Indeed, there are considered punctures on the real axis, up to the full 
$\mathrm{SL(2,\mathbb{R})}$ (see also \cite{Cachazo:2016ror,Mizera:2017cqs}). 

A third model of the hyperbolic plane is the so called \emph{Hyperboloid model}. In a Minkowski space-time $\mathbb{R}^{1,2}$ (with signature $(-,+,+)$), with coordinates $x=(x^0,x^1,x^2)$, consider the upper branch of the hyperboloid $\mathrm{H}$ of equation $x^2=1$. 
We can induce a positive definite metric on it, which turns it into a Riemannian manifold isometric to $(\D,\rho_D)$. We can construct such isometry as follows. Put a disk of euclidean radius 1 at the origin of the spatial plane $x^0=0$, 
and map every point of $\h$ to the disk by a projection through the point $(-1,0,0)$, see fig \ref{fig:iperboloide}.
\begin{figure}[!htbp]
\begin{center}
\begin{tikzpicture}
\label{fig:iperboloide}
\draw (0,5) arc (190:-10:4cm and 1cm);
\filldraw [white] (0,5.2) arc (190:-10:4cm and 1cm) -- cycle;
\draw [thick](3.8,-2) -- (3.8,0);
\draw [thick] (4.358,-0.2) -- (3.8,-2);
\shade [left color=green!50!black] (5.8,0) arc (0:360:2cm and 0.5cm);
\draw (5.8,0) arc (0:360:2cm and 0.5cm);
\draw [>=latex][thick, ->](3.8,0) -- (3.8,7);
\shade[left color=red!50!black, right color=red!50!white] (0,5.2) arc (190:350:4cm and 1cm) .. controls (5.7,2) and (1.9,2) .. (0,5.2) -- cycle;
 \draw (0,5.2) arc (190:350:4cm and 1cm) .. controls (5.7,2) and (1.9,2) .. (0,5.2) -- cycle;
 \draw (0,5.2) arc (190:-10:4cm and 1cm);
\filldraw (5.5,3.5) circle (1pt);
\draw [thick] (5.5, 3.5) -- (4.94,1.7);
\draw (0,5) arc (190:350:4cm and 1cm) -- (3.8,0) -- cycle;
\shade[left color=blue!30!white,right color=blue!5!white,opacity=0.5] (0,5) arc (190:350:4cm and 1cm) -- (3.8,0) -- cycle;
\filldraw [red] (4.358,-0.2) circle (0.9pt);
\draw [thick] (4.94,1.7) -- (4.358,-0.2);
\end{tikzpicture}
\caption{The projection sends a point on the hyperboloid to a point on the disk.}\label{fig:iperboloide}
\end{center}
\end{figure}
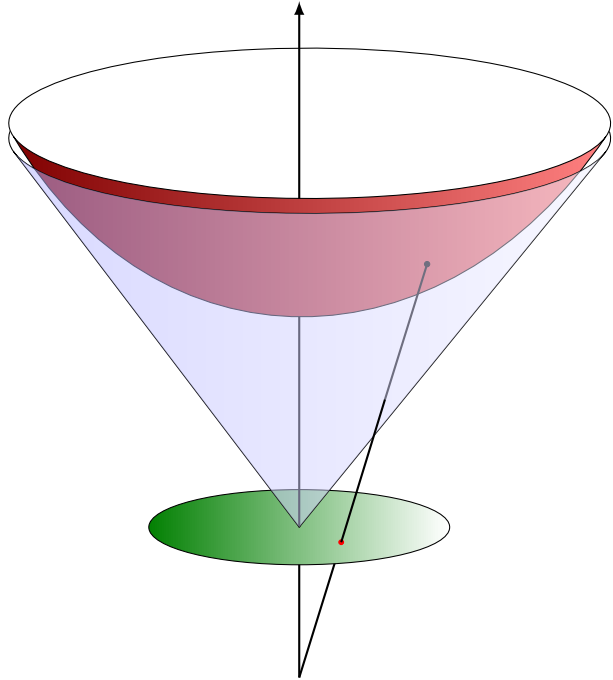
\FloatBarrier
Under this isometry, an element of $SO^\uparrow(1,2)$ corresponds to an element of $\mathrm{Aut}_{\mathbf{D}}$. Using the hyperboloid model, we can compute explicitly the distance between two points $P$ and $Q$ corresponding to time-like 
vectors $l_P$ and $l_Q$, 
using the formula  
\begin{align*}
\cosh(d(P,Q)) = l_P \cdot l_Q. 
\end{align*}
Moreover, many geometrical objects on the hyperbolic plane can be represented by planes in the hyperboloid model: if $k,l$ and $v$ are respectively null, time, and space like momenta, define the planes $w \cdot k = 1,$ $w \cdot v = 1$ and $w \cdot l = 0$, $w\in \mathbb R^{1,2}$. The intersection of these planes with $\h$ yields respectively a horocycle, a circle and a geodesic. Later, we will use this fact to construct meaningful maps from the set of kinematical data of a 
scattering process to the moduli space of Riemann surfaces.

\subsection{Hyperbolic Riemann surfaces}

The hyperbolic plane is not only the simplest surface admitting an hyperbolic metric, but it is the most fundamental one. The Riemann uniformization theorem establishes that, up to bi-holomorphisms, there are only three simply connected Riemann 
surfaces: The Riemann sphere $\mathbb{CP}^1$, the complex plane $\mathbb{C}$ and the Poincar\'e disk $\mathbf{D}$.

A powerful tool in studying Riemann surfaces is the concept of covering space: every Riemann surface $X$ admits a universal covering surface $\tilde{X}$ which is simply connected, therefore, must be one among the three above. It turns out that almost every 
surface is covered by $\mathbf{D}$, the only exceptions being $\mathbb{C}$, $\mathbb{CP}^1$ and a torus without punctures. The crucial property of the universal covering space is that it allows to lift any continuous map $f : X \to X'$ to a  map of 
covering spaces $\tilde{f}: \tilde{X} \to \tilde{X'}$. In particular, we can lift loops on $X$, thought of as maps $\gamma : [0,1] \to X$ with $\gamma(0) = \gamma(1)$. Because it is a loop, its lift $\tilde{\gamma}$ is a path that connects preimages of the 
base point $\gamma(0)$, however $\tilde{\gamma}$ does not have to be a loop itself, and indeed it is not unless $\gamma$ is contractible to a point. If $\gamma$ is not trivial as a loop, $\tilde{\gamma}$ can be used to define an element of 
$\mathrm{Aut}_{\mathbf{D}}$. This produces a representation of the fundamental group $\pi(X)$ with values in $\mathrm{Aut}_{\mathbf{D}}$, the image of which is a torsion-free discrete subgroup of $\mathrm{Aut}_{\mathbf{D}}$, these groups are called 
\emph{Fuchsian groups}. The knowledge of the covering map $\mathbf{D} \to X$ yields a Fuchsian group, but we can go in the opposite direction: for a given Fuchsian group $\Gamma$ the quotient $\mathbf{D}/\Gamma$ yields a Riemann surface. 

With the above construction, we can equip (most) Riemann surfaces with an hyperbolic metric, because we can think of a small neighbourhood $U \subset X$ as bi-holomorphic to a subset of $\mathbf{D}$ and therefore induce the metric $\rho_\mathbf{D}$ in 
$U$. This is another way to characterise the complex structure of $X$: $X$ and $X'$ are isometric if and only if they are bi-holomorphic. 

As mentioned in the introduction, one of the most important results of this approach is that we can uniquely decompose a Riemann surface in fundamental building blocks, called pair of pants. The complex structure of $X$ is then encoded in the pattern of 
gluings and in the Fenchel-Nielsen coordinates, which give a real-analytic structure to Teichm\"uller space.

\subsection{Bordered Riemann surfaces}

The pair of pants decomposition can be extended to Riemann surfaces with boundary \cite{abikoff, devadoss}.
A bordered Riemann surface is defined to be a two dimensional orientable manifold with boundary, together with an holomorphic atlas which induces an analytic structure on the boundary. In the study of their moduli we have to add $m$ punctures and $n$ markings, which are respectively marked points in the interior and in the various boundary components of the surface, in order to maintain stability. The space of these surfaces up to automorphisms is denoted $\mathcal{M}_{g,h;m,n}$ where $g$ is the genus, $h$ the number of bordered components, $m$ and $n$ the number of markings on the interior and on the boundary (our notation is a bit simplified with respect to that of \cite{devadoss}).

Any (bordered with markings) Riemann surface $X$ admits a unique hyperbolic metric with the property that every marked interior point becomes a cusp, every marked boundary point an half cusp and every ordinary point in the boundary has a 
neighbourhood isometric to a neighbourhood of a purely imaginary number in $\mathbb{H} \cap \{\mathrm{Re}(z) \ge 0 \}$, this is called the \emph{intrinsic} metric of $X$. To find this metric, we construct a mirror version of $X$: this is done by considering 
a copy of the same underlying topological surface and covering it with the same atlas used for $X$, however now we say that a point has complex coordinate $\bar{z}$ rather than $z$. The two coordinates match on a boundary component so we can glue $X$ and its mirror to create a unique surface, \emph{the double} $X^\mathbb{C}$.
$X^\mathbb{C}$ has no boundary and admits an hyperbolic metric so that the seams of this gluing are geodetical: the restriction of this metric yields the intrinsic metric of $X$.

The usual pair of pants construction is then generalised so that the pants decomposition of $X^\mathbb{C}$ is compatible with the gluing. In $X$ this yields geodetical arcs and loops whose lengths are the coordinates in the moduli space. The details of this construction can be found in \cite{abikoff,devadoss}.

In the following we will consider two specific cases: The disk with $n$ markings on its boundary, and the annulus with  $n$ markings on the outer boundary. Their doubles produce a Riemann sphere with punctures on a circle and a torus with markings on a cycle. 

\section{Tree Level}
\label{sec:tree}

In this section we focus on the tree level amplitudes of the bi-adjoint scalar field theory relative to a single order, that is of the form $m(\alpha|\alpha)$ according to the notation used in literature. We will begin with a study of the moduli space $\mathcal{M}_{\mathbf{D},n}$ 
of the disk with markings on the boundary, which is known to be an Associahedron. We will see how the hyperboloid model suggests the existence of a natural \emph{kinematical} Associahedron, which is different from the one proposed in \cite{Arkani-Hamed:2017mur}. Finally we will comment on the problem of extracting the amplitude from the canonical form of our realisation of the kinematical Associahedron.

\subsection{The moduli space Associahedron}

We now unwind the hyperbolic approach applied to the moduli space of disks with $n$ markings on the boundary, which we call $\mathcal{M}_{\mathbf{D},n}$ for brevity. Notice that the complex double of such a disk is a punctured Riemann sphere, with punctures on a circle. Therefore, it is natural to expect a connection with the real section of the moduli space of genus 0 Riemann surfaces, where the Associahedron studied in \cite{Arkani-Hamed:2017mur} belongs.

\begin{figure}[!htbp]
\begin{center}
\begin{tikzpicture}[>=latex,scale=0.75, every node/.style={transform shape}]
\filldraw[cyan!30!white] (0,0) circle (2.4cm);
\draw [ultra thick] (0,0) circle (2.4cm);

\foreach \x/\y in {1.158/-2.136,2.319/-0.7425,2.19/1.064,-1.384/2.006,-2.416/0.2381}
\shade [ball color=red] (\x,\y) circle (4pt);
\draw[red] (-0.05461,-2.394) -- (0.08732,2.393);
\draw[line width=4pt,->] (2.4+0.4,0) -- (2.4+0.2+1+0.5,0);
\begin{scope}[xshift=4.8cm+2cm]
\filldraw[cyan!30!white] (0,0) circle (2.4cm);
\draw [ultra thick] (0,0) circle (2.4cm);

\foreach \x/\y in {-1.384/2.006,-2.416/0.2381}
\shade [ball color=red] (\x,\y) circle (4pt);
\shade [ball color=red](2.4,0) circle (4pt);

\end{scope}

\begin{scope}[xshift=4.8cm+2cm+4.8cm]
\filldraw[cyan!30!white] (0,0) circle (2.4cm);
\draw [ultra thick] (0,0) circle (2.4cm);
\foreach \x/\y in {1.158/-2.136,2.319/-0.7425,2.19/1.064}
\shade [ball color=red] (\x,\y) circle (4pt);
\shade [ball color=red](-2.4,0) circle (4pt);

\end{scope}

\end{tikzpicture}
\caption{Contraction of a geodesic arc produce two nodal disks.}
\label{fig:arccontraction}
\end{center}
\end{figure}
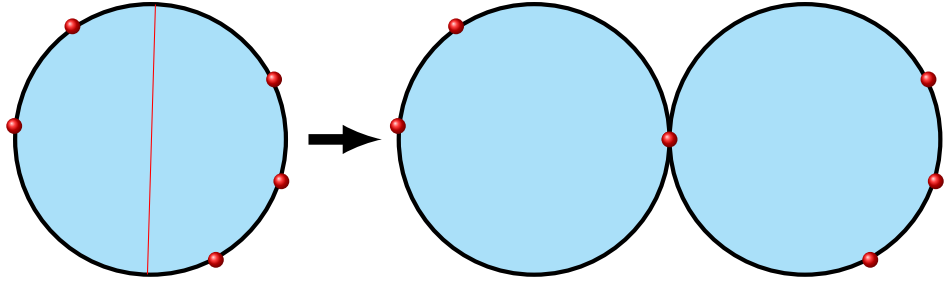

Let $X$ be a disk with markings and $X^\mathbb{C}$ the sphere resulting from the doubling procedure. Any pants decomposition of $X^\mathbb{C}$ consists of geodesics that suitably partition the punctures: on the original copy of $X$ these are geodesic arcs that touch non adjacent boundary components. 
We can consider an $n$-gon obtained substituting punctures and geodesic boundaries with edges and vertices. Then, it is then obvious that the geodesic arcs are in bijection with the diagonals of the $n$-gon: this establishes the combinatorial equivalence of 
$\mathcal{M}_{\mathbf{D},n}$ with the Associahedron $K_{n-1}$. 
As shown in figure \ref{fig:arccontraction}, by contracting a diagonal we get two nodal disks and therefore we recover the usual factorisation $K_{n-1} \approx K_{m} \times K_{n-m}$.
\FloatBarrier

By an explicit realisation of $X^\mathbb{C} = \mathbf{D}/\Gamma$ one can compute the modulus carried by a diagonal in $X$. 
However, in the simple case at hand we can proceed as follows.
Let $X = (\mathbf{D}; z_1 \dots z_n),$ by the Riemann mapping theorem $X$ is bi-holomorphic to the geodetical convex hull of the points $z_1 ,\dots, z_n$, therefore we can study directly the moduli of $\mathrm{Conv}(z_1, \dots, z_n)$ for which we 
already have the intrinsic metric: it is the restriction of the metric $\rho_\mathbf{D}$ of the Poincar\'e disk. 

Let $\alpha$ be a diagonal separating the punctures in two sets  $z_L = \{z_j,z_{j+1},\dots , z_{i}\}$ and $z_R = \{z_{i+1},z_{i+2}, \dots , z_{j-1}\}$, as in figure \ref{fig:diagonalicross}. 
\begin{figure}[!htbp]
\begin{center}
\begin{tikzpicture}[>=latex,scale=0.65, every node/.style={transform shape}]
\node at (-6cm,0) {};
\filldraw[cyan!30!white] (0,0) circle (4cm);
\draw [ultra thick] (0,0) circle (4cm);
\draw [green!60!black, ultra thick] (1.3,2.56) .. controls (0.6,1.2) and (0,-1.5) .. (0,-2.6);
\draw [blue, ultra thick] (-0.5,3.96863) .. controls (0.4,2.5) and (2,2.2) .. (3,2.64575) .. controls (2.8,2) and (3.3,0.5) .. (4,0) .. controls (2.4,-0.7) and (2.1,-2.8) .. (2.32,-3.12) .. controls (1.5,-2.4) and (-1.5,-2.4) .. (-2.4,-3.2) .. controls (-2.2,-2.4) and (-3.2,-1) .. 
(-3.9, -0.888819) .. controls (-3,0) and (-2.8,1.3) .. (-3.2,2.4) .. controls (-2.4,2.6) and (-1.3,3) .. (-0.5,3.96863);
\shade [ball color=red] (-0.5,3.96863) circle (4pt);
\shade [ball color=red] (3,2.64575) circle (4pt);
\shade [ball color=red] (4,0) circle (4pt);
\shade [ball color=red] (2.4,-3.2) circle (4pt);
\shade [ball color=red] (-2.4,-3.2) circle (4pt);
\shade [ball color=red] (-3.9, -0.888819) circle (4pt);
\shade [ball color=red] (-3.2,2.4) circle (4pt);
\node at (3.3,3+0.2) {$\pmb {j-1}$};
\node at (-0.5,4.3+0.2) {$\pmb j$};
\node at (2.6,-3.45-0.2) {$\pmb {i+1}$};
\node at (-2.5,-3.4-0.2) {$\pmb {i}$};
\node at (0,0) {$\pmb {\alpha}$};
\node (a) at (3.9,1.5) {};
\node (b) at (6,2) {};
\draw[blue] [-latex, bend left, ultra thick] (a) to (b);
\node at (7.3,2) {$\pmb {(j, i+1|i,j-1)}$};
\end{tikzpicture}
\caption{The convex hull of several points is shown, the blue lines are its geodesic boundaries. The green line is the geodesic arc with the minimal length among those separating the particles in this way.}
\label{fig:diagonalicross}
\end{center}
\end{figure}
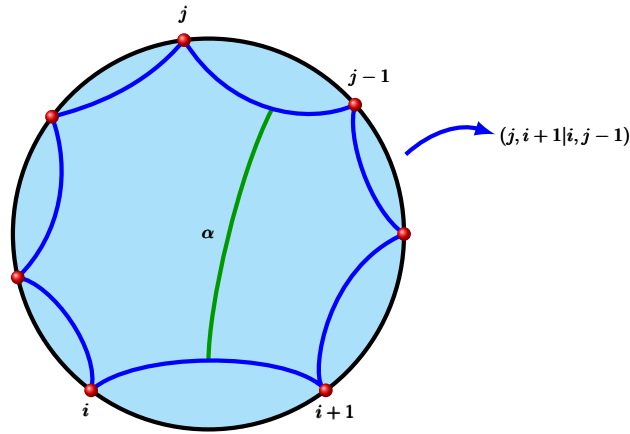
\FloatBarrier
We have to find the shortest geodetic in the homotopy class of $\alpha$ (t. i. separating the punctures in the same sets): first perform a Moebius map to send $(z_{i+1},z_{j-1},z_{j},z_{i})$ to $(-b,-a,a,b)$ on the real axis, then the geodetic in question (which we still call $\alpha$) is the line $\mathrm{Re}(\sigma) = 0$, and the geodesic arc has length $\mathrm{arcosh}(1+\frac{(b-a)^2}{2a b}).$ We can write this result in a more appealing form as 
\begin{align*}
  l_\alpha = \mathrm{arcosh}(1 - 2(j\
  (i+1)|i\ (j-1)),
\end{align*}
where
\begin{align*}
  (i j | k l) = \frac{(z_k - z_i)(z_l - z_j)}{(z_k - z_j)(z_l - z_i)},
\end{align*}
is a cross ratio. It is evident that we can trade the geodesic lengths for the cross ratios as moduli for $X$. Geodesic lengths are obviously positive and are bounded by inequalities. This translates in having the cross ratios negative and 
bounded by inequalities which guarantee that the ciclic order of the particles is preserved.  Notice that the cross ratios we consider here are not the same as those considered in \cite{Arkani-Hamed:2017mur}.
We recall that cross ratios satisfy relations which we can summarise as
\begin{align} 
  (ab|cd) &= 1 - (ac|bd), \cr
  (ab|cd) &= \frac{1}{(ba|cd)}, \label{eq:crrelations} \\
  (ab|cd) &= (ab|ce)(ab|ed) \quad (\text{cocycle condition}).\nonumber
\end{align}

Suppose that $\alpha$ is being contracted, then either the punctures $\sigma_j, \sigma_{i}$ or the punctures $\sigma_{i+1},\sigma_{j-1}$ are collapsing and, because of the cyclic order, this forces the punctures between them to collapse as well. 
However this does not mean that the lengths of the remaining diagonals have to vanish, neither on the left nor on the right side of $\alpha$, only that if we want to compute them in terms of puncture variables then we have to introduce a new set of variables for the nodal disks.

\subsection{The kinematical Associahedron}

Let us begin with the most natural definition of the kinematical space, namely the set of momenta $(k_1,\dots,k_n),$ with $\sum_{a=1}^n k_a = 0$ and $k_a^2 = 0$, up to a common Lorentz transformation. This space has natural boundaries, given by the 
singular regions where some Mandelstam variable vanishes, that is where $k_I^2 = 0$ with $k_I = \sum_{a \in I} k_a$.

In view of the hyperboloid model, this kinematical space naturally projects to $\mathcal{M}_{\mathbf{D},n}$. The map is constructed as follows: let $l_i$ be a collection of time-like vectors such that $l_i \to k \in L^+,$ then the points $l_i/\sqrt{l_i^2}$ on the hyperboloid are 
mapped by the projection discussed in section \ref{sec:review} to a sequence of points that converge to $z \in \partial \mathbf{D},$ put $\phi(k) := z$. Then, we can take our collection of momenta $(k_1, \dots, k_n)$ and send them to a collection of points 
$(z_1,\dots,z_n)$ on $\partial \mathbf{D}$ (the ones with negative energy are first mapped to $-k_a$). Note that another set of momenta $k'_a = \Lambda k_a$, with $\Lambda \in \mathrm{SO}^{\uparrow}(1,2)$ is sent to $(z_1', \dots, z_n')$ with $z' = \gamma(z)$ for a 
suitable $\gamma \in \mathrm{Aut}_\mathbf{D}$. Therefore, we defined a map from the kinematical space to $\mathcal{M}_{\mathbf{D},n}$. 

However, note that both $k_a$ and $\lambda k_a$ are sent to the same point in $\partial{D}$: the map $\phi$, being projective in nature, forgets about the scale of the momenta. Because of this, we re-define our kinematical space to be up to single rescaling of 
the momenta \footnote{When this does not create confusion we will simply write $k_a$ for its light ray $\{ \lambda k\ | \lambda \in \mathbb{R}\}$}. To be honest, there is no physical reason to do this from the point of view of amplitudes, quite the contrary it will be the root of the problems in extracting the amplitude.

Note that the map $\phi$ does not depend on the momentum conservation requirement, so we can relax this assumption: any set of $n$ null-momenta is a viable choice to represent $n$ light rays.

In conclusion, we have defined the kinematical space $\mathcal{K}_{n}$ to be the set of $n$ distinct light rays up to $\mathrm{SO}^{\uparrow}(1,2)$, and we described a map $\phi$
\begin{align}
  \phi\ :\ &\mathcal{K}_{n} \to \mathcal{M}_{\mathbf{D},n}\\ \nonumber
           &(k_1, \dots, k_n) \mapsto X=(\mathbf{D};\phi(k_1),\dots, \phi(k_n)).
\end{align}

Before going further, let us stress that our choice of kinematical space differs from the one made in \cite{Arkani-Hamed:2017mur}. There the kinematical space is defined directly by the Mandelstam variables $s_{I} = (\sum_{i \in I} k_i)^2$. It is therefore a vector space obtained solving the momentum conservation constraints and virtually independent on the dimensionality $D$ of the space time. However, for a specific $D$, Mandelstam variables have to satisfy other non linear constraints. This fact forces us to consider the momenta $k$ 
themselves, on the other hand this is a very natural choice from the point of view of the hyperboloid model.

Notice that $\mathcal{K}_n$ is not a compact space, since we removed the configurations of collinear light rays. We can compactify this space in a method virtually identical to the moduli space compactification in terms of punctures. Suppose some set 
of momenta $k_{i \in L}$ is becoming collinear, then move to the reference frame comoving with $k_L =\sum_{i \in L} k_i$. In this frame the momenta $k_j \in R = L^c$ may or not stay at fixed positions as $k_L^2 = 0$. To represent all possible situations we use a diagonal 
in the Poincar\'e disk: a diagonal separating $k_a$ into subsets $L$ and $R$ represents a limit where $k_L^2 = 0$ and the particles in $R$ fail to remain at finite positions in the center of mass frame of $k_L$ but rather collapse to a unique light ray. If we stick in the frame comoving with 
$K_R$ instead, we see the particles in $L$ becoming collinear, and thus we get a natural factorisation of the kinematical space in $\mathcal{K}_{|L|+1} \times \mathcal{K}_{|R|+1}$. Therefore the compactification of the kinematical space is naturally identified 
with the Associahedron $K_{n-1}$. It is not difficult to see that $\phi$ extends to a continuous map from the kinematical Associahedron to the moduli space Associahedron.

We now relate the map $\phi$ with the scattering equations. We can choose a specific Lorentz frame and write a null momentum $k$
\begin{align*}
k= \left(\begin{array}{c} \mathcal{E} \\ \mathcal{E} \vec{\eta} \end{array} \right)
\end{align*}
where $\vec{\eta} = (\cos(\theta),\sin(\theta))$ is a unit norm vector and $\mathcal{E}$ may be positive or negative. Such $k$ is sent to $z = \exp(i \theta)$, so we can write the Mandelstam variables as 
\begin{align}
\label{eq:kijzizj}
k_a \cdot k_b = \mathcal{E}_a \mathcal{E}_b (1-\cos(\theta_a -\theta_b)) = \frac 12 \mathcal{E}_a \mathcal{E}_b |z_a - z_b|^2.
\end{align}
The scattering equations for the kinematics $k_a$ and their images $z_a$ in the disk now are
\begin{align}
E_a = \sum_{b \ne a} \frac{k_a \cdot k_b}{z_a - z_b} = \frac 12 \sum_{b \ne a} \mathcal{E}_a \mathcal{E}_b (\bar{z}_a - \bar{z}_b) = -\frac 12 ( \mathcal{E}_a^2 \bar{z}_a - \mathcal{E}_a^2 \bar{z}_a) = 0,
\end{align}
in the last passage we recognised $z_a \mathcal{E}_a$ as the spatial part of $k_a$ and used energy and momentum conservation. Since changing Lorentz frame is equivalent to perform a Moebius map on the punctures $z_a$ and the scattering 
equations are covariant under Moebius map, our initial choice of frame was uninfluential. We conclude that our map lands in a solution of the scattering equations when it acts on the subset of $\mathcal{K}_n$ where momentum conservation holds\footnote{We would like to remark that the solution to the scattering equations provided by the map $\phi$ can be reinterpreted as the well known pair of solutions provided by the spinor helicity formalism \cite{fairlie}.}. However 
we noted that the map $\phi$ is defined without requiring momentum conservation and thus, in a sense, extends the map given by the scattering equations to other regions of the kinematical space. Moreover, it is guaranteed to land in the Associahedron 
$\mathcal{M}_{\mathbf{D},n}$ as long as the momenta $k_a$ are in the cyclic order $(12\dots n)$. Finally, let us remark that for $n>5$ there are in general other solutions to the scattering equations other than the one provided by $\phi$.

This is a good point to make a remark on the peculiarity of $1+2$ dimensional kinematics. Recall from \cite{devadoss} that in $\mathcal{M}_{\mathbf{D},n}$ the moduli carried by diagonals are essentially cross ratios. Using the hyperboloid model, we can quickly express 
the moduli of the surface $X = \phi(k_1,\dots,k_n),$ in terms of the momenta $k_a$. From \eqref{eq:kijzizj} it is immediate to see that
\begin{align}
\label{eq:lcr}
(a b|c d) = \sqrt{\frac{k_{ac}k_{db}}{k_{cb} k_{da}}},
\end{align}
we call the quantity on the right hand side of \eqref{eq:lcr} a \emph{space time cross ratio} because of its similarity to a cross ratio.
Note that the space time cross ratios are invariant under rescaling of each single momentum and is thus a function of the light rays, as they should. Equation \eqref{eq:lcr}
allows us to think of the null momenta $k_a$ as set of \emph{homogeneous coordinates} for the moduli space $\mathcal{M}_{\mathbf{D},n}$.
As we discussed earlier, cross ratios have to satisfy the identities \eqref{eq:crrelations} and because of \eqref{eq:lcr} the space time cross ratios have to satisfy these relations as well, which implies that the Mandelstam variable have to satisfy further relations beyond those coming from momentum conservation. Indeed, for any set of $4$ null momenta, the first of \eqref{eq:crrelations} implies
\begin{align}
  \label{eq:spacetimeplucker}
  \sqrt{k_{12}k_{34}} - \sqrt{k_{13}k_{24}} + \sqrt{k_{14}k_{34}} = 0,
\end{align}
the combinatorics behind this relation is not new in the world of scattering amplitudes, it appears as Pl\"ucker relations in the Grassmannian $\mathrm{Gr}(2,4)$, in the Jacobi identities underlying the colour-kinematics duality and was discovered in 
\cite{Arkani-Hamed:2017mur} to be a condition for the projectivity of a scattering form with numerators. Note that in the particular case where $\sum_{a=1}^4 k_a = 0,$ the condition \eqref{eq:spacetimeplucker} simply reduce to momentum conservation. However, any four momenta 
$k_i$ have to satisfy \eqref{eq:spacetimeplucker} for the mere reason of being null and in $1+2$ dimensions.
Indeed, in three dimensions we can always write
\begin{align*}
\sum_{i=1}^4 c_i k_i = 0,
\end{align*}
let us choose $c_4 = 1,$ then the massless condition implies 
\begin{align}
\label{eq:masslesscondition}
\frac{1}{2} k_4 \cdot k_4 = c_1 c_2 k_{12} + c_1 c_3 k_{13} + c_2 c_3 k_{23} = 0,
\end{align}
but we can also express the coefficients $c_i$ in terms of Mandelstam variables,
\begin{align*}
- k_{14} &= c_2 k_{21} + c_3 k_{13}, \\
- k_{24} &= c_1 k_{14} + c_3 k_{23}, \\
- k_{34} &= c_1 k_{13} + c_2 k_{23},
\end{align*}
and together with the massless condition we get 
\begin{align*}
k_{14} &= \frac{c_2 c_3}{c_1} k_{23},\\
k_{24} &= \frac{c_3 c_1}{c_2} k_{31},\\
k_{34} &= \frac{c_1 c_2}{c_3} k_{12}.
\end{align*}
Using these, we can re-write \eqref{eq:spacetimeplucker} as
\begin{align*}
|k_{12}| \sqrt{\frac{c_1 c_2}{c_3}} + |k_{23}| \sqrt{\frac{c_2 c_3}{c_1}} - |k_{13}| \sqrt{\frac{c_1 c_3}{c_2}} = 0, 
\end{align*}
if we multiply by $\sqrt{c_1 c_2 c_3}$ we see that this is equivalent to \eqref{eq:masslesscondition}, if we correctly take into account the relative order of the momenta.

\subsection{Canonical forms}

Having established that $\mathcal{M}_{\mathbf{D},n}$ is an Associahedron, we can write its canonical form, which is defined to be a differential form with logarithmic singularities at the boundaries of $\mathcal{M}_{\mathbf{D},n}$ (also called dlog for brevity). When dealing 
with a simple polytope the canonical form is obtained by the formula \cite{Arkani-Hamed:2017tmz}
\begin{align*}
\omega = \sum_v {\rm sign}(v) \bigwedge_{v \in F} \frac{dF}{F},
\end{align*}
the sum is over all vertices, and the wedge product is taken over all facets defined by the equation $F=0$ incident to that vertex. 

Each vertex of $\mathcal{M}_{\mathbf{D},n}$ is labelled by a maximal set of non intersecting diagonals $\alpha =(\alpha_1, ... , \alpha_{n-3}),$ with associated lengths $l_\alpha$ and cross ratios $\chi_{\alpha}$, therefore it contributes to the dlog with 
\begin{align*}
\omega_v = \bigwedge_{i=1}^{n-3} D(\chi_{\alpha_i}),
\end{align*}
where we introduced the shorther notation $D(x) = \frac{dx}{x}$.

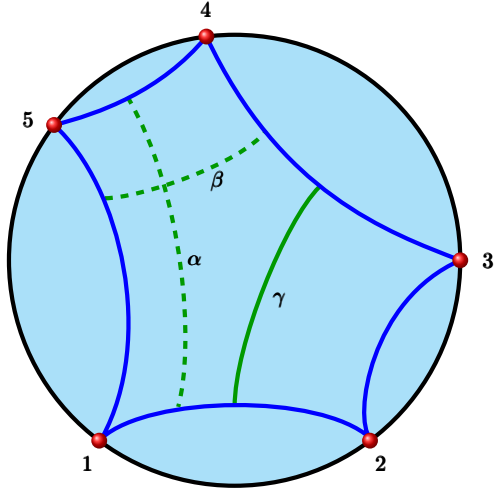
\begin{figure}[!htbp]
\begin{center}
\begin{tikzpicture}[>=latex,scale=0.75, every node/.style={transform shape}]
\filldraw[cyan!30!white] (0,0) circle (4cm);
\draw [ultra thick] (0,0) circle (4cm);
\draw [green!60!black, ultra thick] (1.5,1.3) .. controls (0.9,0.7) and (0,-1.5) .. (0,-2.6);
\draw [green!60!black, ultra thick, dashed] (-1.9,2.9) .. controls (-0.9,1.2) and (-0.7,-1.5) .. (-1,-2.6);
\draw [green!60!black, ultra thick, dashed] (-2.3,1.1) .. controls (-1.7,1.1) and (0.1,1.7) .. (0.56,2.3);
\draw [blue, ultra thick] (-0.5,3.96863) .. controls (0.8,1.2) and (2.7,0.5) .. (4,0) .. controls (2.4,-0.7) and (2.1,-2.8) .. (2.4,-3.15) .. controls (1.5,-2.35) and (-1.5,-2.4) .. (-2.35,-3.15) .. controls (-1.4,-1.4) and (-2,1.3) .. (-3.2,2.4) 
.. controls (-2.4,2.6) and (-1.3,3) .. (-0.5,3.96863);
\shade [ball color=red] (-0.5,3.96863) circle (4pt);
\shade [ball color=red] (4,0) circle (4pt);
\shade [ball color=red] (2.4,-3.2) circle (4pt);
\shade [ball color=red] (-2.4,-3.2) circle (4pt);
\shade [ball color=red] (-3.2,2.4) circle (4pt);
\node at (-2.6,-3.6) {$\pmb 1$};
\node at (2.6,-3.6) {$\pmb 2$};
\node at (4.5,0) {$\pmb 3$};
\node at (-0.5,4.45) {$\pmb 4$};
\node at (-3.65,2.5) {$\pmb 5$};
\node at (-0.7,0) {$\pmb \alpha$};
\node at (-0.3,1.4) {$\pmb \beta$};
\node at (0.8,-0.7) {$\pmb \gamma$};
\end{tikzpicture}
\caption{Three diagonals related by the cocycle condition.}
\label{fig:3diagonali}
\end{center}
\end{figure}

Curiously, it turns out that we only need a single term to find the dlog. At first sight it is not clear how a single set of diagonals can reproduce all the facets of $\mathcal{M}_{\mathbf{D},n}$, but this is a consequence of the cross ratios relations 
\eqref{eq:crrelations}. We can see explicitly how this works for $n=4$ and $n=5$. For $n=4$ there are just two possible diagonals and their cross ratios are related by
\begin{align*}
(42|13) = \frac{1}{(13|24)},
\end{align*}
therefore
\begin{align*}
  D(42|13) = -D(13|24).
\end{align*}
In the $n=5$ case, consider the diagonals represented in figure \ref{fig:3diagonali}.
\FloatBarrier
\noindent
 The cross ratios involved are related by the cocycle condition:
 \begin{align*}
   \chi_\alpha = (52|14) = (14|52) = (14|53)(14|32) = (14|53)(1-(42|13)) = \chi_\beta (1 - \chi_\gamma),
 \end{align*}
therefore
\begin{align*}D (\chi_\alpha) \wedge D(\chi_\gamma) = (D(\chi_\beta) + D(1-\chi_\gamma))\wedge D(\chi_\gamma) = D(\chi_\beta) \wedge D(\chi_\gamma).
\end{align*}

Finally, we can prove that our dlog is the correct one by a direct comparison with the usual Parke-Taylor dlog, 
\begin{align*}
\omega_{PT,n} = \frac{\bigwedge_{i=1}^n d\sigma_i}{\mathrm{Vol}(\mathrm{SL}(2,\mathbb{C}))}\frac{1}{\prod_{i=1}^n(\sigma_i-\sigma_{i+1})} = 
\frac{\bigwedge_{a=4}^{n} d\sigma_a}{\sigma_{45} \sigma_{56} \dots \sigma_{n-1n} \sigma_n}.
\end{align*}
We can canonically define our dlog by using the vertex associated to the diagonals in figure \ref{fig:canonicaldiagonal}, then it reads
\begin{align*}
\omega_n = \bigwedge_{\alpha} D(\chi_\alpha) = \bigwedge_{a=4}^{n} D(a2|1(a-1)).
\end{align*}
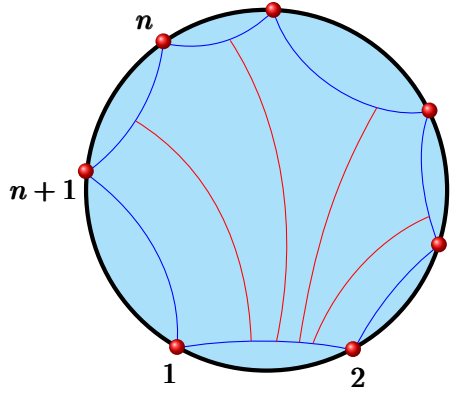
\begin{figure}[!htbp]
\begin{center}
\begin{tikzpicture}[>=latex,scale=1, every node/.style={transform shape}]
\filldraw[cyan!30!white] (0,0) circle (2.4cm);
\draw [ultra thick] (0,0) circle (2.4cm);

\draw[blue] (-1.193,-2.103) .. controls (-0.5799,-1.977) and (0.4932,-1.964) .. (1.162,-2.131) .. controls (1.413,-1.601) and (1.887,-1.03) .. (2.291,-0.7372) .. controls (2.054,-0.1518) and (1.97,0.6007) .. (2.166,1.047) .. controls (1.357, 0.9073) and (0.3399, 1.548) .. (0.08909,2.385) .. controls (-0.2733,2.036) and (-0.7192,1.827) .. (-1.374,1.953) .. controls (-1.444,1.311) and (-1.806,0.6565) .. (-2.406,0.2244) .. controls (-1.597,-0.3469) and (-1.109,-1.225) .. (-1.193,-2.103);

\foreach \x/\y in {-1.191/-2.086,1.149/-2.114,2.291/-0.7213,2.166/1.061,0.09053/2.399,-1.372/1.981,-2.403/0.2537}
\shade [ball color=red] (\x,\y) circle (3pt);

\draw[red] (0.6176, -2.036).. controls (0.9522, -1.171) and (1.594, -0.5857) .. (2.165, -0.3487);
\draw[red] (0.4363, -2.022).. controls (0.6315, -0.767) and (0.8965, 0.09758) .. (1.468, 1.102);
\draw[red] (0.1295, -2.022).. controls (0.4502, -0.516) and (0.1992, 0.9482) .. (-0.4841, 1.994);
\draw[red] (-0.2052, -2.008).. controls (-0.247, -0.7391) and (-0.763, 0.3207) .. (-1.739, 0.9203);

\node at (-1.28,-2.342-0.1) {$\pmb 1$};
\node at (1.223,-2.394-0.1){$\pmb 2$};
\node at (-1.513-0.1,2.135+0.1){$\pmb n$};
\node at (-2.609-0.35,0.2898-0.3){$\pmb {n+1}$};
\end{tikzpicture}
\caption{A canonical choice of diagonals: they all start from the edge $(12)$ and touch the non adjacent edges.}
\label{fig:canonicaldiagonal}
\end{center}
\end{figure}
\FloatBarrier
\noindent Notice that adding a particle simply adds a diagonal and therefore adds a factor $D(n+1\ 2|1n)$ and that in the gauge fixing we used to write $\omega_{PT,n}$, i.e. with $\sigma_1 = 0,\sigma_2 = 1, \sigma_3 = \infty$, we have 
$(n+1\ 2|1n) = \frac{\sigma_{n+1}(\sigma_n-1)}{(\sigma_n-\sigma_{n+1})}$. Suppose by induction that our formula is correct up to $n$, then for $n+1$ we have
\begin{align*}
\omega_{n+1} &= D(n+1\ 2|1n)\wedge \omega_{PT,n} = (D(\sigma_{n+1})-D(\sigma_n-\sigma_{n+1})+D(\sigma_{n}-1)) \wedge \omega_{PT,n} \\
&= \frac{d\sigma_{n+1} \sigma_n}{(\sigma_n - \sigma_{n+1})(\sigma_{n+1})}\wedge \omega_{PT,n} = \omega_{PT,n+1},
\end{align*}
where in the last passage we used the fact that $d\sigma_{n}$ vanish under the wedge with $\omega_{PT,n}$.

What we found so far fits in the general ideas of the Amplituhedra formalism. We have a geometrical object ($\mathcal{M}_{\mathbf{D},n}$) which factorises as an amplitude should. On this space we have a canonical form ($\omega_{PT}$) with logarithmic 
singularities, and finally we a have a map ($\phi$) from kinematics to this object which maps singular regions of the kinematics to the singularities of $\omega_{PT}$. The final step would be to use $\phi$ to find a kinematical avatar of $\omega_{PT}$ from 
which we may read the amplitude, once a canonical measure on the kinematical space has been defined.

\begin{figure}[!htbp]
\begin{center}
\begin{tikzpicture}[>=latex,scale=1, every node/.style={transform shape}]

\begin{scope}[rotate=30]
\draw[ultra thick] (-2.5,0) -- (-1.5,-1) -- (-0.5,0);
\node (k1) at (-2.5,0.2){};
\node (k2) at (-0.5,0.2){};
\node (a) at (-1.5,-1){};
\end{scope}
\begin{scope}[rotate=-30]
\draw[ultra thick] (2.5,0) -- (1.5,-1) -- (0.5,0);
\node (k3) at (2.5,0.2){};
\node (k4) at (0.5,0.2){};
\node (b) at (1.5,-1){};
\end{scope}
\draw[ultra thick] (a.center) -- (b.center) node (c) [midway]{} node(h)[pos = 0.25]{} node(i)[pos = 0.75]{};
\draw[ultra thick] (c.center) --++ (0,-2) node[circle,fill=lightgray,draw,minimum size = 1cm, pin={[thick,pin edge = {ultra thick,black}]-150:$$},pin={[thick,pin edge = {ultra thick,black}]-30:$$},pin={[thick,pin edge = {ultra thick,black}]-90:$$} ] (d) [pos=1]{} node (f)[pos=0.5]{};
\draw[dashed, thick] ($(f.south west) + (-0.5cm,-0.4cm)$) .. controls (f.west) and (f.east) .. ($(f.south east) + (0.5cm,-0.4cm)$);
\draw[dashed, thick] ($(h.south west) + (-0.5cm,-0.4cm)$) .. controls (h.south) and (h.north) .. ($(h.north west) + (0.1cm,0.3cm)$);
\draw[dashed, thick] ($(i.south east) + (0.5cm,-0.4cm)$) .. controls (i.south) and (i.north) .. ($(i.north east) + (-0.1cm,0.3cm)$);

\node at (k1.north) {$k_1$};
\node at (k2.north) {$k_2$};
\node at (k3.north) {$k_4$};
\node at (k4.north) {$k_3$};


\end{tikzpicture}
\caption{The cutted propagators of this Feynman diagram are on-shell, this together with momentum conservation at the cubic vertices forces $k_1$ and $k_3$ to be proportional.}
\label{fig:feynfail}
\end{center}
\end{figure}
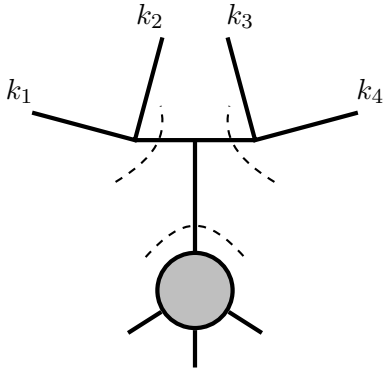

Since the amplitude is expressed as a rational function on the propagators, it is tempting to try to use the lasts as coordinates on the kinematical space and therefore express the measure as something like $d^{n-3} k_{ij}$. However the single Mandelstam
variables are not invariant under rescaling, and if we want to use them as coordinates defined on $\mathcal{K}_n$ we have to perform a gauge fixing and impose some of them to be constants. Of course, this is very reminiscent of the constraints imposed 
in \cite{Arkani-Hamed:2017mur} on the Mandelstam variables.
For $n=4$, with momentum conservation enforced, this works perfectly fine: we can impose $k_{13} \doteq c_{13}$ and use either $k_{12}$ or $k_{14}$ to be a coordinate. $\omega_{PT}$ is easily pushed to the canonical form of $\mathcal{K}_{n}$:
\begin{align*}
\omega_{\mathcal{K}_n} := \phi_*\omega_{PT} = \phi_* D(42|13) = D\left(\sqrt{\frac{k_{14} k_{32}}{ k_{12} k_{34}}}\right) = D\left(\frac{k_{14}}{k_{12}}\right) = dk_{14} \left(\frac{1}{k_{14}} + \frac{1}{k_{12}}\right),
\end{align*}
and we can extract the amplitude $\mathcal{A}_4$. Unfortunately, this approach does not work anymore from $n=5$ onward. One may think that this is due to the fact that we cannot simply pushforward the form as above, but rather we have to sum over all the 
solutions of the scattering equations. However for $n=5$ this is not a possible explanation, because the solution given by $\phi$ is the unique one. The problem is deeper and the solution harder: any gauge fixing $k_{ij} \doteq c_{ij}$ fails to see some of 
the boundaries of the kinematical Associahedron. Indeed, in order to have a multi-particle propagator $k_{I} = 0,$ in $D=3$ we need to make all the particles involved collinear\footnote{This is not strictly true for a single propagator, but if we want to make all the propagators of a Feynman diagram on-shell then we are forced to break some constraint}, and therefore eventually we will break the gauge $k_{ij} \doteq c_{ij}$. 
Moreover, one cannot hope to solve this problem simply by increasing $D$: consider a Feynman diagram such as the one in figure \ref{fig:feynfail}. If the cut propagators are on-shell, momentum conservation at the cubic vertices forces $k_1$ and $k_3$ to 
become collinear for any $D$, and therefore $k_{13}=0$ as well.
\FloatBarrier
\noindent This is not really in contrast with the approach of \cite{Arkani-Hamed:2017mur}, because there, mandelstam variables are thought of as ``abstract variables'', not necessarily coming from a scalar product 
of two D-dimensional momenta $k_{ij} = k_i \cdot k_j$.

Finally, recall that the reason why we defined the kinematical space only on the light-rays was that $\phi$ forgets the scale of the momenta $k_a$. Whilst this is natural from the point of view of the hyperboloid model, we already noted that it is a bit 
awkward from the point of view of the amplitude. Indeed, the tree level amplitude $\mathcal{A}_n$ transforms well under a global rescaling $k_i \to \lambda k_i$ and, therefore, it make sense to couple it with a factor $d^{n-3} k_{ij}$ to make an invariant 
object. On the other hand $\mathcal{A}_n$ does not have any simple behaviour under independent rescaling $k_i \to \lambda_i k_i$ of the single momenta, so that there is really no hope that $\omega_{\mathcal{K}_n}$ may be written as $d^{n-3}k\ \mathcal{A}_n$.

We postpone to the final section a discussion on how these problems may be solved.

\section{Loop level}
\label{sec:loop}
We now extend the construction described in the last section to include loop momenta. Mimicking the tree level case, we will begin discussing the hyperbolic approach to moduli space of Riemann surfaces of genus 1. In the same way as the 
Associahedron emerged from the diagonals of the $n$-gon, the geodesic lengths on a $n$-gon with an extra boundary component realise a polytope called Halohedron, which has many promising features to be identified with the 1-loop Amplituhedron for 
the cubic theory. We will then generalise the kinematical space to include a loop momentum and accordingly extend the map $\phi$ to this kinematical space. Similarly to what happened at tree level, this will lead to a compactification of the kinematical 
space which is immediately recognised as an avatar of the Halohedron in space time. However, this time we will not compute the moduli in a conformal fashion nor even try to write the dlog form or the amplitude. Again, we postpone discussing a possible
solution to this problem to the final section.

\subsection{The moduli space $\mathcal{M}_{\mathbf{D},1;n}$}

Consider a disk with $n$ markings on the outer boundary and a disk removed from the interior: the doubling of this surface produces a torus. We then tentatively take as definition for a real section on the moduli space of marked tori the surfaces we obtain 
by this doubling, and we move to the study of $\mathcal{M}_{\mathbf{D},1;n}$.

As described in \cite{devadoss} it turns out that $\mathcal{M}_{\mathbf{D},1;n}$ is a polytope, called \emph{Halohedron} and denoted $H_n$, which is a close kin of the Associahedron. Indeed, both the Halohedron and the Associahedron fit in a very general combinatorial 
description in terms of marked tubings on a graph. This picture also includes the Cyclohedron $W_n$ whose poset of facets represents the possible ways to associate particles on a circle rather than on a segment like the Associahedron.

\begin{figure}[!htbp]
\begin{center}
\begin{tikzpicture}[>=latex,scale=0.5, every node/.style={transform shape}]
\shade [left color=green!40!white, right color=green!10!white] (0,10) -- (3.464105, 8) -- (4.33013, 6.5) -- (4.33013, 2.5) -- (0.866025, 0.5) -- (-0.866025, 0.5) -- (-4.33013, 2.5) -- (-4.33013, 6.5) -- (-3.464105, 8) -- cycle;
\shade [left color=green!60!white, right color=green!40!white] (-0.866025, 6.5) -- (-3.464105, 8) -- (-4.33013, 6.5) -- (-1.73205, 5) -- cycle;
\shade [left color=green!40!white, right color=green!20!white] (0.866025, 6.5) -- (3.464105, 8) -- (4.33013, 6.5) -- (1.73205, 5) -- cycle;
\shade [left color=green!50!white, right color=green!30!white] (0.866025, 0.5) -- (-0.866025, 0.5) -- (-0.866025, 3.5) -- (0.866025, 3.5) -- cycle;
\shade [left color=green!70!white, right color=green!40!white] (0.866025, 6.5) -- (-0.866025, 6.5) -- (-1.73205, 5) -- (-0.866025, 3.5) -- (0.866025, 3.5) -- (1.73205, 5) -- cycle;
\draw [ultra thick] (0,10) -- (3.464105, 8) -- (4.33013, 6.5) -- (4.33013, 2.5) -- (0.866025, 0.5) -- (-0.866025, 0.5) -- (-4.33013, 2.5) -- (-4.33013, 6.5) -- (-3.464105, 8) -- cycle;
\draw [ultra thick] (-0.866025, 6.5) -- (-3.464105, 8) -- (-4.33013, 6.5) -- (-1.73205, 5) -- cycle;
\draw [ultra thick] (0.866025, 6.5) -- (3.464105, 8) -- (4.33013, 6.5) -- (1.73205, 5) -- cycle;
\draw [ultra thick] (0.866025, 0.5) -- (-0.866025, 0.5) -- (-0.866025, 3.5) -- (0.866025, 3.5) -- cycle;
\draw [ultra thick] (0.866025, 6.5) -- (-0.866025, 6.5) -- (-1.73205, 5) -- (-0.866025, 3.5) -- (0.866025, 3.5) -- (1.73205, 5) -- cycle;
\draw [dashed,opacity=0.5] (-4.33013, 2.5) -- (0,5); 
\draw [dashed,opacity=0.5] (4.33013, 2.5) -- (0,5); 
\draw [dashed,opacity=0.5] (0, 10) -- (0,5); 
\shade [ball color=red] (0,10) circle (3pt);
\shade [ball color=red]  (3.464105, 8) circle (3pt);
\shade [ball color=red] (4.33013, 6.5) circle (3pt);
\shade [ball color=red] (4.33013, 2.5) circle (3pt);
\shade [ball color=red] (0.866025, 0.5) circle (3pt);
\shade [ball color=red] (-0.866025, 0.5) circle (3pt);
\shade [ball color=red] (-4.33013, 2.5) circle (3pt);
\shade [ball color=red] (-4.33013, 6.5) circle (3pt);
\shade [ball color=red] (-3.464105, 8) circle (3pt);
\shade [ball color=red, opacity=0.5] (0, 5) circle (3pt);
\shade [ball color=red] (0.866025, 6.5) circle (3pt);
\shade [ball color=red] (-0.866025, 6.5) circle (3pt);
\shade [ball color=red] (-1.73205, 5) circle (3pt);
\shade [ball color=red] (1.73205, 5) circle (3pt);
\shade [ball color=red] (-0.866025, 3.5) circle (3pt);
\shade [ball color=red] (0.866025, 3.5) circle (3pt);
\filldraw [cyan!30!white] (7,8) circle (1.2cm);
\filldraw [white] (6.5,8) circle (0.4cm);
\draw [ultra thick, red] (5.9,8.5) .. controls (7.5,8.9) and (7.5,7.1) .. (5.9,7.5);
\draw [ultra thick] (7,8) circle (1.2cm);
\draw [ultra thick] (6.5,8) circle (0.4cm);
\shade [ball color=red] (7,9.2) circle (4pt);
\node at (7,9.5) {\pmb {3}};
\shade [ball color=red] (7,6.8) circle (4pt);
\node at (7,6.5) {\pmb {1}};
\shade [ball color=red] (8.2,8) circle (4pt);
\node at (8.5,8) {\pmb {2}};
\filldraw [cyan!30!white] (8.5,5) circle (1.2cm);
\filldraw [white] (8,5) circle (0.4cm);
\draw [ultra thick, red] (8.9,6.1) -- (8.9,3.9);
\draw [ultra thick] (8.5,5) circle (1.2cm);
\draw [ultra thick] (8,5) circle (0.4cm);
\shade [ball color=red] (7.3,5) circle (4pt);
\node at (7.1,5) {\pmb {1}};
\shade [ball color=red] (9.3,5.9) circle (4pt);
\node at (9.3,6.2) {\pmb {3}};
\shade [ball color=red] (9.3,4.1) circle (4pt);
\node at (9.3,3.8) {\pmb {2}};
\filldraw [cyan!30!white] (6.6,2) circle (1.2cm);
\filldraw [white] (6.6,2) circle (0.4cm);
\draw [ultra thick, red] (6.6,1.6) -- (6.6,0.8);
\draw [ultra thick] (6.6,2) circle (1.2cm);
\draw [ultra thick] (6.6,2) circle (0.4cm);
\shade [ball color=red] (6.6,3.2) circle (4pt);
\node at (6.6,3.5) {\pmb {3}};
\shade [ball color=red] (5.4,2) circle (4pt);
\node at (5.2,2) {\pmb {1}};
\shade [ball color=red] (7.8,2) circle (4pt);
\node at (8.1,2) {\pmb {2}};
\node (a) at (7.4,4.7) {};
\node (b) at (3.5,4) {};
\draw[blue] [-latex, bend left, ultra thick] (a) to (b);
\node (c) at (5.8,8.2) {};
\node (d) at (3.5,7) {};
\draw[blue] [-latex, bend right, ultra thick] (c) to (d);
\node (e) at (5.4,1.8) {};
\node (f) at (2.5,1.5) {};
\draw[blue] [-latex, bend left, ultra thick] (e) to (f);
\node (g) at (-3.5,9.5){};
\node (h) at (0,5.8) {};
\draw[blue] [-latex, bend left, ultra thick] (g) to (h);
\filldraw [cyan!30!white] (-5,9.5) circle (1.2cm);
\filldraw [white] (-5,9.5) circle (0.4cm);
\draw [ultra thick] (-5,9.5) circle (1.2cm);
\draw [ultra thick, red] (-5,9.5) circle (0.4cm);
\shade [ball color=red] (-5,10.7) circle (4pt);
\shade [ball color=red] (-5-1.2,9.5) circle (4pt);
\shade [ball color=red] (-5+1.2,9.5) circle (4pt);
\node (i) at (-5+0.2,10.7+0.3) {\pmb {3}};
\node (n) at (-5-1.2-0.2,9.5+0.2){\pmb {1}};
\node (m) at (-5+1.2+0.3,9.5+0.2){\pmb {2}};
\end{tikzpicture}
\caption{The three dimensional Halohedron. There are four ``tadpole factorisation'' square facets, three ``factorisation'' pentagons, three ``cut'' pentagons and a Cyclohedron. The Cyclohedral and tadpole factorisation facets acts as a ``bridge'' pairing the tadpoles vertices of the physical facets. }
\label{fig:h3diagonali}
\end{center}
\end{figure}
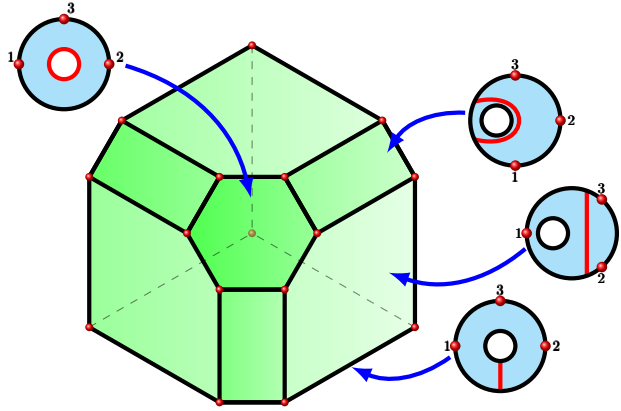
\FloatBarrier

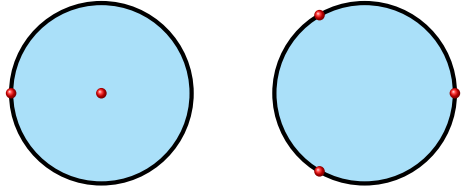
\begin{figure}[!htbp]
\begin{center}
\begin{tikzpicture}[>=latex,scale=0.5, every node/.style={transform shape}]
\filldraw[cyan!30!white] (0,0) circle (2.4cm);
\draw [ultra thick] (0,0) circle (2.4cm);
\shade [ball color=red] (-2.4,0) circle (4pt);
\shade [ball color=red] (0,0) circle (4pt);

\begin{scope}[xshift=7cm]
\filldraw[cyan!30!white] (0,0) circle (2.4cm);
\draw [ultra thick] (0,0) circle (2.4cm);
\shade [ball color=red] (0:2.4) circle (4pt);
\shade [ball color=red] (120:2.4) circle (4pt);
\shade [ball color=red] (-120:2.4) circle (4pt);
\end{scope}

\end{tikzpicture}
\caption{The smallest possible stable components that appear when contracting arcs.}
\label{fig:smallest}
\end{center}
\end{figure}

The description of $H_n$ relevant to scattering amplitudes is in terms of arcs on a marked annulus. The arcs are obtained by the pair of pants decomposition of the torus and unlike in the prievous section can now be of different types. The arcs have again an associated modulus, the length of the geodesic they represent, if the modulus is zero the arc is contracted and we obtain nodal disks. As for $\mathcal{M}_{\mathbf{D},n}$, only those arcs that upon contraction produce stable nodal components are admissible, but now we have a new kind of stable lowest dimensional component: a disk with a marked point in the interior and a marked point on the boundary, see figure \ref{fig:smallest}.
Contracting an arc produce a codimension one facet of $H_n$.

\FloatBarrier

As proved in \cite{devadoss}, the facets of $H_n$ are exactly one Cyclohedron $W_n$, $n$ Associahedra $K_{n+1}$ and $n^2-n$ facets of the form $K_m \times  H_{n-m+1}$ for $n \ge m \ge 2$.
In table \ref{tab:facethaloedron} are shown the possible arcs of the Halohedron and the corresponding contractions, in figure \ref{fig:h3diagonali} the three dimensional Halohedron $H_3$ is shown with some of its facets labelled.

Some of the facets of $H_n$ have an immediate physical interpretation. Factorisation facets with $m<n$ are obviously in 1-1 correspondence with the possible factorisations of a 1-loop planar integrand, and the factorisation $K_m \times  H_{n-m+1}$ mirrors the factorisation of the integrand in a lower point integrand and tree level amplitude. Similarly, we have an Associahedral facet for each possible cut of the integrand, and since Associahedra compute tree level amplitudes, we can interpret these facets as reflecting the forward limit of a cutted integrand.

\begin{table}
\centering
\begin{tabular}{|c|l|}
\hline
\parbox[c]{0.4\linewidth}{
\begin{tikzpicture}[>=latex,scale=0.6, every node/.style={transform shape}]
\filldraw [cyan!30!white] (0,0) circle (2.4cm);
\filldraw [white] (0,0) circle (0.8cm);
\draw [ultra thick] (0,0) circle (2.4cm);
\draw [ultra thick,red] (0,0) circle (0.8cm);
\foreach \x/\y in {-2.124/1.061,-1.414/1.939,0.5641/2.343,1.901/1.479,2.263/-0.7909,1.386/-1.975,-1.525/-1.863}
{ \shade [ball color=red] (\x,\y) circle (4pt);}
\draw [line width=4pt,->] (2.8,0) -- (4,0);
\node at (-2.124-0.4,1.061) {$\pmb 1$};
\node at (-1.525-0.4,-1.863) {$\pmb n$};

\begin{scope}[xshift=7cm]
\filldraw [cyan!30!white] (0,0) circle (2.4cm);
\draw [ultra thick] (0,0) circle (2.4cm);
\foreach \x/\y in {-2.124/1.061,-1.414/1.939,0.5641/2.343,1.901/1.479,2.263/-0.7909,1.386/-1.975,-1.525/-1.863}
{ \shade [ball color=red] (\x,\y) circle (4pt);}
\node at (-2.124-0.4,1.061) {$\pmb 1$};
\node at (-1.525-0.4,-1.863) {$\pmb n$};
\shade [ball color=red] (0,0) circle (4pt);
\end{scope}

\node at (0,2.4+0.2){};
\node at (0,-2.4-0.2){};

\end{tikzpicture}
} & Cyclohedron $W_n$
\\
\hline
\parbox[c]{0.6\linewidth}{
\begin{tikzpicture}[>=latex,scale=0.6, every node/.style={transform shape}]

\filldraw [cyan!30!white] (0,0) circle (2.4cm);
\filldraw [white] (-1.2,0) circle (0.5cm);
\draw [ultra thick] (0,0) circle (2.4cm);
\draw [ultra thick] (-1.2,0) circle (0.5cm);
\draw [red, ultra thick] (-2,1.3) .. controls (-0.0198, 0.8665) and (-0.0198, -0.8665) .. (-2,-1.3);
\shade [ball color=red] (2.4,0) circle (4pt);
\foreach \x/\y in {0.6755/2.301,2.124/1.117,1.776/-1.627,-0.578/-2.323}
{\shade [ball color=red](\x,\y) circle (4pt);}

\node at (-1.386, 1.953+0.4){$\pmb i+1$};
\node at (-1.386, -1.953-0.4){$\pmb i$};
\shade [ball color=red] (-1.386, 1.953) circle (4pt);
\shade [ball color=red] (-1.386, -1.953) circle (4pt);

\draw [line width=4pt,->] (2.8,0) -- (4,0);

\begin{scope}[xshift=7cm]
\filldraw [cyan!30!white] (0,0) circle (2.4cm);
\draw [ultra thick] (0,0) circle (2.4cm);
\shade [ball color=red] (2.4,0) circle (4pt);
\filldraw [white] (0,0) circle (0.8cm);
\draw [ultra thick] (0,0) circle (0.8cm);
\end{scope}
\begin{scope}[xshift=9.4cm+2.4cm]
\filldraw [cyan!30!white] (0,0) circle (2.4cm);
\draw [ultra thick] (0,0) circle (2.4cm);
\node at (-1.386, 1.953+0.4){$\pmb i+1$};
\node at (-1.386, -1.953-0.4){$\pmb i$};
\shade [ball color=red] (-1.386, 1.953) circle (4pt);
\shade [ball color=red] (-1.386, -1.953) circle (4pt);
\shade [ball color=red] (-2.4,0) circle (4pt);
\foreach \x/\y in {0.6755/2.301,2.124/1.117,1.776/-1.627,-0.578/-2.323,2.4/0}
{\shade [ball color=red](\x,\y) circle (4pt);}
\end{scope}

\end{tikzpicture}

} & Tadpole Factorisation $K_n \times H_1$
\\
\hline
\parbox[c]{0.55\linewidth}{
\begin{tikzpicture}[>=latex,scale=0.6, every node/.style={transform shape}]
\filldraw [cyan!30!white] (0,0) circle (2.4cm);
\filldraw [white] (0,0) circle (0.8cm);
\draw [ultra thick, red] (-1.3,2) -- (-1.3,-2);
\draw [ultra thick] (0,0) circle (2.4cm);
\draw [ultra thick] (0,0) circle (0.8cm);
\foreach \x/\y in {-1.79/1.591,-2.18/1.006,-2.375/0.2954,-2.166/-1.014,-1.706/-1.696,-0.2577/-2.365,1.4/-1.961,2.403/-0.09453,1.957/1.382,-0.2994/2.371}
{
\shade [ball color=red] (\x,\y) circle (4pt);
}
\draw [line width=4pt,->] (2.8,0) -- (4,0);
\node at (-1.789, 1.757+0.3) {$\pmb i$};
\node at (-1.789, -1.757-0.3) {$\pmb j$};
\node at (-0.2981, 2.371+0.3) {$\pmb i+1$};
\node at (-0.2981, -2.371-0.3) {$\pmb j-1$};
\begin{scope}[xshift=7cm]
\filldraw [cyan!30!white] (0,0) circle (2.4cm);
\draw [ultra thick] (0,0) circle (2.4cm);
\shade [ball color=red] (2.4,0) circle (4pt);
\foreach \x/\y in {-1.706/-1.699,-2.18/-1.002,-2.361/0.2941,-2.18/1.005,-1.775/1.59}
{
\shade [ball color=red] (\x,\y) circle (4pt);
}
\node at (-1.789, 1.757+0.3) {$\pmb i$};
\node at (-1.789, -1.757-0.3) {$\pmb j$};
\end{scope}
\begin{scope}[xshift=9.4cm+2.4cm]
\filldraw [cyan!30!white] (0,0) circle (2.4cm);
\draw [ultra thick] (0,0) circle (2.4cm);
\filldraw [white] (0,0) circle (0.8cm);
\draw [ultra thick] (0,0) circle (0.8cm);

\node at (-1.386, 1.953+0.4){$\pmb i+1$};
\node at (-1.386, -1.953-0.4){$\pmb j-1$};
\shade [ball color=red] (-1.386, 1.953) circle (4pt);
\shade [ball color=red] (-1.386, -1.953) circle (4pt);
\shade [ball color=red] (-2.4,0) circle (4pt);
\foreach \x/\y in {0.6755/2.301,2.124/1.117,1.776/-1.627,-0.578/-2.323,2.4/0}
{\shade [ball color=red](\x,\y) circle (4pt);}

\end{scope}

\end{tikzpicture}
} & Factorisation $K_{m} \times H_{n-m+1}$
\\
\hline
\parbox[c]{0.75\linewidth}{
\begin{tikzpicture}[>=latex,scale=0.6, every node/.style={transform shape}]
\filldraw[cyan!30!white] (0,0) circle (2.4cm);
\filldraw[white] (0,0) circle (0.8cm);
\draw [ultra thick] (0,0) circle (2.4cm);
\draw [ultra thick] (0,0) circle (0.8cm);
\shade [ball color=red] (2.4,0) circle (4pt);
\shade [ball color=red] (-2,1.3) circle (4pt);
\shade [ball color=red] (-2,-1.3) circle (4pt);
\foreach \x/\y in {-1.595/1.786,0.6755/2.301,2.124/1.117,1.776/-1.627,-0.578/-2.323}
{\shade [ball color=red](\x,\y) circle (4pt);}
\draw [red, ultra thick] (-2.4,0) -- (-0.8,0);
\draw [line width=4pt,->] (2.8,0) -- (4,0);
\node at (-2.2-0.3,1.5) {$\pmb i+1$};
\node at (-2.2,-1.5) {$\pmb i$};

\begin{scope}[xshift=7cm]
\filldraw[cyan!30!white] (0,0) circle (2.4cm);
\filldraw[white] (-1.6,0) circle (0.8cm);
\draw [ultra thick] (-1.6,0) circle (0.8cm);
\draw [ultra thick] (0,0) circle (2.4cm);
\shade [ball color=red] (2.4,0) circle (4pt);
\shade [ball color=red] (-2,1.3) circle (4pt);
\shade [ball color=red] (-2,-1.3) circle (4pt);
\draw [line width=4pt,->] (2.8,0) -- (4,0);
\foreach \x/\y in {-1.595/1.786,0.6755/2.301,2.124/1.117,1.776/-1.627,-0.578/-2.323}
{\shade [ball color=red](\x,\y) circle (4pt);}
\node at (-2.2-0.3,1.5) {$\pmb i+1$};
\node at (-2.2,-1.5) {$\pmb i$};

\end{scope}
\begin{scope}[xshift=14cm]
\filldraw [cyan!30!white] (0,0) circle (2.4cm);
\draw [ultra thick] (0,0) circle (2.4cm);
\shade [ball color=red] (2.4,0) circle (4pt);
\shade [ball color=red] (-2,1.3) circle (4pt);
\shade [ball color=red] (-2,-1.3) circle (4pt);
\shade [ball color=red] (1.386, 1.953) circle (4pt);
\shade [ball color=red] (1.386, -1.953) circle (4pt);
\node at (-1.985-0.4, 1.298){$\pmb +$};
\node at (-1.985-0.4, -1.298){$\pmb -$};
\node at (1.386+0.2, 1.953+0.4){$\pmb i+1$};
\node at (1.386, -1.953-0.4){$\pmb i$};
\foreach \x/\y in {2.11/1.159,2.333/0.6297,2.389/-0.3592,2.221/-0.9441}{\shade [ball color=red](\x,\y) circle (4pt);}
\end{scope}
\end{tikzpicture}

} & Cut Associahedron $K_{n+1}$
\\
\hline
\end{tabular}
\caption{Realisation of different faces of $H_n$.}
\label{tab:facethaloedron}
\end{table}
\FloatBarrier

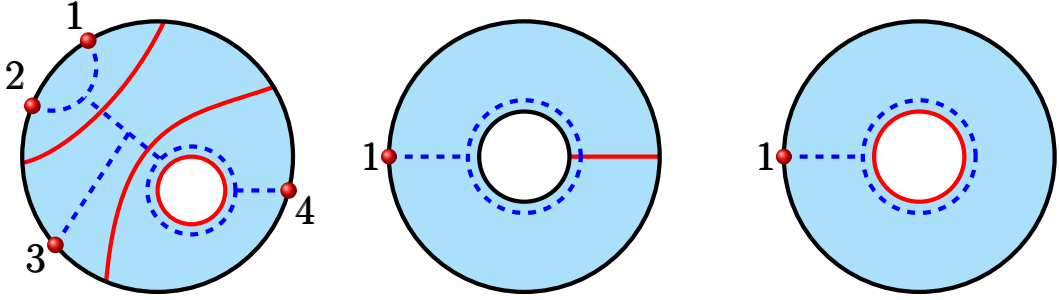
\begin{figure}[!htbp]
\begin{center}
\begin{tikzpicture}[>=latex,scale=0.75, every node/.style={transform shape}]

\begin{scope}[scale=0.6,every node/.style={transform shape}]
\filldraw [cyan!30!white] (0,0) circle (4cm);
\draw [red, ultra thick]   (-3.98,-0.18) .. controls (-3,0) and (-1,1.5) ..  (0.19,3.99);

\filldraw [white] (1,-1) circle (1cm);
\draw [red, ultra thick] (-1.52,-3.7).. controls (-1.1,1.2) and (1,1.2) .. (31:4);
\draw [blue,dashed,ultra thick](-3.7,1.5) arc (243:400:1.3);
\draw [ultra thick] (0,0) circle (4cm);
\draw [red,ultra thick] (1,-1) circle (1cm);
\draw [blue,ultra thick,dashed] (1,-1) circle (1.3cm);
\draw [blue,ultra thick,dashed] (0.05,-0.05) -- (-2.15,1.75);
\draw [blue,ultra thick,dashed] (-0.85,0.7) -- (-3.03,-2.63);
\draw [blue,ultra thick,dashed] (2.3,-1) -- (3.9,-1);
\shade [ball color=red] (-3.7,1.5) circle (7pt);
\shade [ball color=red] (-2.05,3.43) circle (7pt);
\shade [ball color=red] (3.87,-1) circle (7pt);
\shade [ball color=red] (-3.02,-2.61) circle (7pt);
\end{scope}
\node at (120:2.9) {\huge $\pmb 1$}; 
\node at (150:2.9) {\huge $\pmb 2$}; 
\node at (220:2.8) {\huge $\pmb 3$}; 
\node at (-20:2.8) {\huge $\pmb 4$};

\begin{scope}[xshift=10cm]
\filldraw [cyan!30!white] (-3.5,0) circle (2.4cm);
\filldraw [white] (-3.5,0) circle (0.8cm);
\draw [ultra thick, red] (-2.7,0) -- (-1.1,0);
\draw [ultra thick, dashed, blue] (-3.5,0) circle (1cm);
\draw [ultra thick, dashed, blue] (-5.9,0) -- (-4.5,0);
\draw [ultra thick] (-3.5,0) circle (2.4cm);
\draw [ultra thick] (-3.5,0) circle (0.8cm);
\shade [ball color=red] (-5.9,0) circle (4pt);
\node at (-6.2,0) {\huge $\pmb 1$};
\filldraw [cyan!30!white] (3.5,0) circle (2.4cm);
\filldraw [white] (3.5,0) circle (0.8cm);
\draw [ultra thick, dashed, blue] (3.5,0) circle (1cm);
\draw [ultra thick, dashed, blue] (1.1,0) -- (2.5,0);
\draw [ultra thick] (3.5,0) circle (2.4cm);
\draw [ultra thick, red] (3.5,0) circle (0.8cm);
\shade [ball color=red](1.1,0) circle (4pt);
\node at (0.9-0.1,0) {\huge $\pmb 1$};
\end{scope}

\end{tikzpicture}
\caption{Examples of Feynman diagrams dual to choices of arcs. The middle and right figures show how tadpoles are created, we call these ``IR'' and ``UV'' tadpole respectively. }
\label{fig:esempifeyn}
\end{center}
\end{figure}

\begin{figure}[!htbp]
\begin{center}
\begin{tikzpicture}[>=latex,scale=0.6, every node/.style={transform shape}]
\filldraw [green!20!white] (0,0) -- (3,0) -- (3,-6) -- (-3,-6) -- (-3,-3) -- cycle;
\draw [ultra thick] (0,0) -- (3,0) -- (3,-6) -- (-3,-6) -- (-3,-3) -- cycle;
\shade [ball color =red] (0,0) circle (4pt);
\shade [ball color =red] (3,0) circle (4pt);
\shade [ball color =red] (3,-6) circle (4pt);
\shade [ball color =red] (-3,-6) circle (4pt);
\shade [ball color =red] (-3,-3) circle (4pt);
\draw [ultra thick] (4,-0.2) -- (4,1.2);
\draw [ultra thick] (4,0.5) -- (4.5,0.5);
\draw [ultra thick] (5,0.5) circle (0.5cm);
\draw [ultra thick] (-2,0.3) -- (-2,1.7);
\draw [ultra thick] (-2,1) -- (-1.5,1);
\filldraw [ultra thick] (-1,1) circle (0.5cm);
\draw [ultra thick] (-4,-2.7) -- (-4,-1.3);
\draw [ultra thick] (-4,-2) -- (-4.5,-2);
\filldraw [ultra thick] (-5,-2) circle (0.5cm);
\draw [ultra thick] (-4,-7.7) -- (-4,-6.3);
\draw [ultra thick] (-4,-7) -- (-4.5,-7);
\draw [ultra thick] (-5,-7) circle (0.5cm);
\draw [ultra thick] (4.5,-6.5) -- (4.5,-6);
\draw [ultra thick] (4.5,-7.5) -- (4.5,-8);
\draw [ultra thick] (4.5,-7) circle (0.5cm);
\draw [black,ultra thick, <->] (-3+0.5,-6+0.5) -- (3-0.5,0-0.5);
\draw [black,ultra thick, <->] (-3+0.2,-3) .. controls (-1.419, -2.561) and (-0.09575, -1.133) .. (0,-0.2);

\begin{scope}[xshift=5cm,yshift=-2.8cm]
\begin{scope}[scale=0.5]
\filldraw [cyan!30!white] (0,0) circle (2.4cm);
\filldraw [white] (0,0) circle (0.8cm);
\draw [ultra thick, red] (-2.7+3.5,0) -- (-1.1+3.5,0);
\draw [ultra thick] (0,0) circle (2.4);
\draw [ultra thick] (0,0) circle (0.8);
\shade [ball color=red] (0,-2.4) circle (8pt);
\shade [ball color=red] (0,2.4) circle (8pt);
\end{scope}
\node at (0,-2.4/2-0.4) {\huge $\pmb 1$};
\node at (0,+2.4/2+0.5) {\huge $\pmb 2$};
\end{scope}

\begin{scope}[xshift=0cm,yshift=-8.3cm]
\begin{scope}[scale=0.5,xscale=-1]
\filldraw [cyan!30!white] (0,0) circle (2.4cm);
\filldraw [white] (0,0) circle (0.8cm);
\draw [ultra thick, red] (-2.7+3.5,0) -- (-1.1+3.5,0);
\draw [ultra thick] (0,0) circle (2.4);
\draw [ultra thick] (0,0) circle (0.8);
\shade [ball color=red] (0,-2.4) circle (8pt);
\shade [ball color=red] (0,2.4) circle (8pt);
\end{scope}
\node at (0,-2.4/2-0.4) {\huge $\pmb 1$};
\node at (0,+2.4/2+0.5) {\huge $\pmb 2$};
\end{scope}

\end{tikzpicture}
\caption{$H_2$ is shown with all of its vertices labelled by Feynman diagrams. Note that tadpoles appear in IR-UV pairs (white-black), but also in pairs dictated by the cyclohedron (black arrows). The associahedral facets correspond to the cyclic order $\alpha=(-12+)$ and $\alpha=(-21+)$ in the formula \eqref{eq:songformula}, but they do not intersect at their tadpole vertices.}
\label{fig:h2feynmanass}
\end{center}
\end{figure}
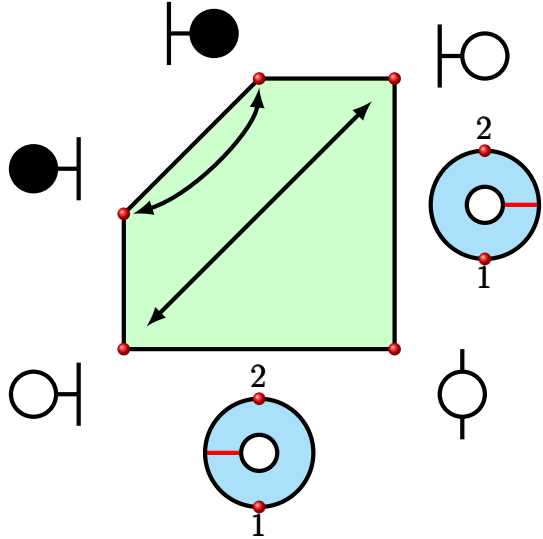

\begin{figure}[!htbp]
\begin{center}
\begin{tikzpicture}[>=latex,scale=0.6, every node/.style={transform shape}]
\begin{scope}[xshift=0cm]
\filldraw[cyan!30!white] (0,0) circle (2.4cm);
\draw [ultra thick] (0,0) circle (2.4cm);
\filldraw[white] (0,0) circle (0.8cm);
\draw [ultra thick] (0,0) circle (0.8cm);
\shade [ball color=red] (2.4,0) circle (4pt);
\draw [ultra thick, dashed, blue] (0,0) circle (1cm);
\draw [ultra thick, dashed, blue] (1,0) -- (2.4,0);
\end{scope}
\begin{scope}[xshift=2.4cm+2.4cm]
\filldraw[cyan!30!white] (0,0) circle (2.4cm);
\draw [ultra thick] (0,0) circle (2.4cm);
\foreach \x/\y in {-1.4/1.953,-0.6476/2.315,0.09053/2.385,1.247/2.05,1.483/-1.891,0.4248/-2.351,-0.578/-2.323,-1.302/-2.003}
{\shade [ball color=red](\x,\y) circle (4pt);}
\draw[red, line width=1pt] (-1.776,1.605) .. controls (-0.9123,1.326) and (1.065,1.382) .. (1.692,1.702);
\draw[red, line width=1pt] (-1.776,-1.605) .. controls (-0.9123,-1.326) and (1.065,-1.382) .. (1.692,-1.702);
\draw [ultra thick, dashed, blue] (0,0) -- (-2.4,0);
\draw [ultra thick, dashed, blue] (0,0) -- (0,1.45);
\draw [ultra thick, dashed, blue] (0,0) -- (0,-1.45);
\shade [ball color=red] (0,1.45) circle(4pt);
\shade [ball color=red] (0,-1.45) circle(4pt);
\shade [ball color=red] (-2.4,0) circle(4pt);
\node at (-1.4,1.953+0.6) {\huge $\pmb {i}$};
\node at (1.247,2.05+0.6) {\huge $\pmb j$};
\node at ($(-1.4-0.1,-1.953-0.4)+(-0.3,-0.3)$) {\huge $\pmb{i-1}$};
\node at ($(1.547,-2.05-0.3)+(0.3,-0.3)$) {\huge $\pmb {j+1}$};

\end{scope}
\draw [line width=2pt,<->] (9.6-2.2,0) -- (9.6-0.4,0);
\begin{scope}[xshift=9.6cm+2.4cm]
\filldraw[cyan!30!white] (0,0) circle (2.4cm);
\draw [ultra thick] (0,0) circle (2.4cm);
\foreach \x/\y in {-1.4/1.953,-0.6476/2.315,0.09053/2.385,1.247/2.05,1.483/-1.891,0.4248/-2.351,-0.578/-2.323,-1.302/-2.003}
{\shade [ball color=red](\x,\y) circle (4pt);}
\draw[red, line width=1pt] (-1.776,1.605) .. controls (-0.9123,1.326) and (1.065,1.382) .. (1.692,1.702);
\draw[red, line width=1pt] (-1.776,-1.605) .. controls (-0.9123,-1.326) and (1.065,-1.382) .. (1.692,-1.702);
\draw [ultra thick, dashed, blue] (0,0) -- (2.4,0);
\draw [ultra thick, dashed, blue] (0,0) -- (0,1.45);
\draw [ultra thick, dashed, blue] (0,0) -- (0,-1.45);
\shade [ball color=red] (0,1.45) circle(4pt);
\shade [ball color=red] (0,-1.45) circle(4pt);
\node at (-1.4,1.953+0.6) {\huge $\pmb {i}$};
\node at (1.247,2.05+0.6) {\huge $\pmb j$};
\node at ($(-1.4-0.1,-1.953-0.4)+(-0.3,-0.3)$) {\huge $\pmb{i-1}$};
\node at ($(1.547,-2.05-0.3)+(0.3,-0.3)$) {\huge $\pmb {j+1}$};


\end{scope}
\begin{scope}[xshift=9.6cm+2.4cm+2.4cm+2.4cm,xscale=-1]
\filldraw[cyan!30!white] (0,0) circle (2.4cm);
\draw [ultra thick] (0,0) circle (2.4cm);
\filldraw[white] (0,0) circle (0.8cm);
\draw [ultra thick] (0,0) circle (0.8cm);
\shade [ball color=red] (2.4,0) circle (4pt);
\draw [ultra thick, dashed, blue] (0,0) circle (1cm);
\draw [ultra thick, dashed, blue] (1,0) -- (2.4,0);
\end{scope}

\end{tikzpicture}
\caption{A vertex of the Cyclohedron is represented by a nodal disk as in the left figure. We can pair it with the vertex represented by the nodal disk in the right.}
\label{fig:tadpolepairing}
\end{center}
\end{figure}
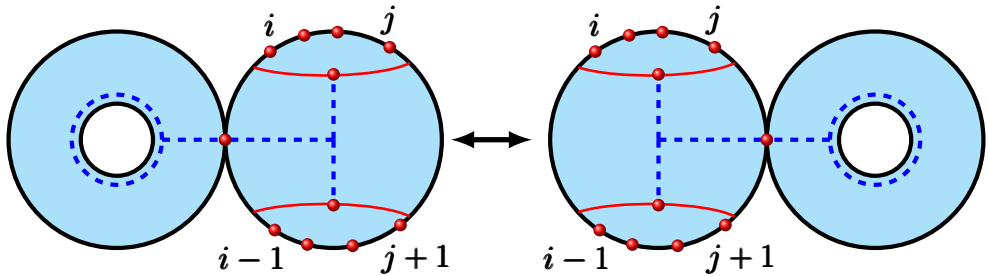

On the other hand, there is no obvious intepretation for the cyclohedral facet and the ``tadpole'' factorisations (those with $m=n$). However, we can find one by looking at the vertices of $H_n$. A vertex corresponds to a maximal choice of arcs. It is easy to see that we can pick a maximum of $n$ non intersecting admissible arcs. To a maximal choice of arcs we can associate a cubic Feynman diagram, essentially in the same way as was done in \cite{Arkani-Hamed:2017mur}: the arcs partition the annulus in zones, to each zone we associate a cubic vertex and then we contract the vertices with propagators intersecting the arcs. Obviously, we now get cubic 1-loop planar diagrams, possibly with bubbles and tadpoles, see figure \ref{fig:esempifeyn}.
\FloatBarrier
Using this rule we can label all the vertices of $H_n$ with Feynman diagrams, this is done explicitly for $H_2$ in figure \ref{fig:h2feynmanass}. As it is clear from figure \ref{fig:esempifeyn}, tadpoles always appear in pairs that we called ``IR-UV'' for reasons that will be explained later. 
\FloatBarrier
Note that the Cyclohedral face has only tadpole vertices. Moreover, when constructing a maximal set of arcs for a vertex of $W_n$ we are basically building a maximal set of arcs for the tree level Associahedra with $n+1$ particles, $\mathcal{M}_{\mathbf{D},n+1}$. The tadpole is inserted in a cubic vertex and, therefore, we can unambiguosly perform the move depicted in figure \ref{fig:tadpolepairing}, the result is that we paired different vertices of $W_n$.
\FloatBarrier
\noindent Because of the ``IR-UV'' pairing, we can extend this cyclohedral pairing to all the tadpole vertices of $H_n$. The picture obtained is strongly reminiscent of a formula proposed by He et al \cite{He:2015yua} that builds the 1-loop integrands of the bi-adjoint theory from the forward limit of tree level amplitudes. We report this formula here
\begin{align}
\label{eq:songformula}
m^{1-loop}_n (\pi|\rho) = \int \frac{d^D l}{(2 \pi)^D} \frac{1}{l^2} \lim_{k_\pm \to \pm l} \sum_{\substack{\alpha \in \mathrm{cyc}(\pi) \\ \beta \in \mathrm{cyc}(\rho)}} m_{n+2}(-\alpha+|-\beta+),
\end{align}
the inner sum is done over all cyclic permutations of the order $\pi$ and $\rho$ and $m(-\alpha+|-\beta+)$ is the tree level double-partial amplitude for $n+2$ particles, where the $+2$ stands for particles $\pm$ that carry momenta $k_{\pm}$ in the forward limit. Computing these tree level amplitudes as a sum over Feynman diagrams one gets cutted tadpole diagrams, but these come in pairs with opposite signs and cancel. In particular, in our case with  $\rho = \pi = (1,\dots,n)$, for fixed $\alpha$ the sum over $\beta$ produces the amplitude $m_{n+2}(- \alpha +|-\alpha +)$ plus the off diagonal elements that cancel the tadpoles of $m_{n+2}(- \alpha +|-\alpha +)$. The cancellation happens following the Cyclohedral pairing we just saw. Thanks to the recent interpretation of the tree level scattering amplitudes as intersection numbers of associahedra in $\mathcal{M}_n(\mathbb{R})$ we can interpret this fact as due to the particular tiling of $\mathcal{M}_n(\mathbb{R})$ in Associahedra and Permutohedra. In our picture though, the Associahedra do not intersect at their tadpole vertices, and the pairing happens thanks to the extra facets $W_n$ and $K_n \times H_1$. In a sense then, these extra facets are there to act as a ``bridge'' that connects different Associahedra in such a way that their tadpoles are paired in the same cancelling pattern discovered by He et al.

To summarise, in this section we saw that the moduli space $\mathcal{M}_{\mathbf{D},1;n}$ is naturally identified with the Halohedron $H_n$. We discovered that its geometry encodes the factorisation and cut properties of a $1$-loop double-partial integrand $m^{\mathrm{1-loop}}(1,\dots,n|1,\dots,n)$ and, in particular, reproduces the cancellation of tadpoles of its forward limit.

\subsection{The kinematical Halohedron and the map $\phi$ at 1-loop}

We understood that the Halohedron encodes several key properties of the 1-loop planar integrands, and has thus the first ingredient to be identified as the 1-loop Amplituhedron for the bi-adjoint theory. The next step of the Amplituhedra formalism, is to find a map from the 
1-loop kinematical space to the Halohedron, which must send singular regions of the former to facets of the latter.

We begin defining the $1$-loop kinematical space $\mathcal{K}_{1,n}$. In virtue of what we saw at tree level, we define it to be the set of $n$ light rays with homogeneous coordinates $k_a$, together with a loop momentum $l$. We will restrict $l$ to be 
time-like and bounded in the region $0< l^2 < 1$. Finally, all is taken up to Lorentz transformations. The time-likeness of $l$ has really no natural justification for now a part the connection with the hyperboloid model. On the other hand, the requirement 
$l^2 < 1$ can be interpreted as a sort of ``cut-off'' that we need to regularise the divergence in the diagrams. Note that the only divergencies we may have in a cubic scalar field theory in $1+2$ dimensions at 1-loop come from the integration of the tadpoles. 
Of course we know that tadpoles cancel and so we do not really need a cut-off, but the Halohedron does not know that and, probably, a cut-off is needed to cancel the pair of UV tadpoles. Indeed this explanation is heavily hand-waving but it seems quite 
plausible.

Having defined our kinematical space, we now need to associate to the kinematical data $(k_1,\dots,k_n;l)$ a unique bordered Riemann surface $X = \phi(k_1,\dots,k_n;l)$. The time momenta has a very naive avatar in the Poincar\'e disk: its projection. 
But this construction is too simple, because projecting we loose any information about its mass, and it does not yield a bordered Riemann surface. Another meaningful geometrical object associated to $l$ is a circle, defined by
\begin{align*}\mathrm{C}_l = \{ w \in \mathrm{H} |\ l \cdot w = 1\}.
\end{align*}
It is easy to see that this is indeed an hyperbolic circle of hyperbolic radius $\cosh(r) = \frac{1}{\sqrt{l^2}}$. 
We can now can define a bordered Riemann surface simply cutting away this circle from the Poincar\'e disk. Of course to the null momenta we keep associating their light-rays and thus the corresponding points in the boundary of $\mathbf{D}$. In conclusion, we have defined a map
\begin{align*}
\phi\ :\ &\mathcal{K}_{1,n} \to \mathcal{M}_{\mathcal{D},1;n} \\
       &(k_1,\dots,k_n;l) \mapsto \mathbf{D}\setminus C_l.
\end{align*}
Note that if the kinematical data $(k_a;l)$ are related to another set of kinematical data by an element $\eta$ of $\mathrm{SO}^\uparrow(1,2)$, the associated surfaces are bi-holomorphic. Indeed the transformation of $\mathrm{Aut}_\mathbf{D}$ corresponding to $\eta$ 
realises the bi-holomorphism. Interestingly the little group of $l$ translates to the automorphism group of the surface $X$: we can picture the surface $X = \phi(k_1,\dots,k_n;l)$ in the frame comoving with $l$, where it looks like an annulus, the little group of $l$ are now rotations around the origin and these are exactly the automorphisms of the annulus.
The crucial point of the map $\phi$ is that when $l^2 \to 0$ the circle $C_l$ becomes a horocycle $H_l$ \footnote{Recall from section \ref{sec:review} that a horocycle is defined by the condition $H_l := \{ w \in \mathrm{H}\ |\ l \cdot w = 1\}$ with $l \in \mathrm{L}^+$}, and thus by normalisation of the surface $\mathbf{D} \setminus H_l$ we obtain a disk with two extra punctures, thus mimicking a forward limit of a cutted integrand.

We now turn to the compactification of the kinematical space. A natural boundary occurs when $l^2 = 1$, that is when $l$ hits the cut-off. In this case the circle $C_l$ shrinks to a point and we get a natural extension of the map $\phi$ that sends this boundary 
to the Cyclohedral face of the Halohedron $H_n$ in moduli space. Recall that the Cyclohedron paired tadpoles in ``IR-UV'' pairs, our nomenclature came from the avatar of the cyclohedral facet in space time.

The next really new ingredient is that the time-like momentum $l$ can become massless or can become asymptotical to a light ray. We have to add meaningful boundaries to $\mathcal{K}_{1,n}$ reflecting these limits. In space time picture, the redundance 
$\mathrm{\Aut}_\mathbf{D}$ is tantamount to choose a Lorentz frame and we now have a new natural choice: the frame comoving with $l$.
Choosing this frame is equivalent to find the element $\Lambda_l \in \mathrm{SO}^{\uparrow}(1,2)$ such that
\begin{align*}
\Lambda_l l = \left( \begin{array}{c} \sqrt{l^2} \\ 0 \\ 0 \end{array} \right).
\end{align*}
This condition fixes $\Lambda_l$ up to elements of the little group of $l$. Suppose now that $l \to p \in \mathrm{L}^+,$ in the topology of $\mathbb{R}^{1,2}$, and without loss of generality suppose that the light ray of $p$ is between particles $i$ and $i+1$ with respect to the planar ordering. We 
can define a canonical $\Lambda_l$ by
\begin{align*}
\Lambda_l := \eta_l \circ \gamma_1 \circ R_{p,1},
\end{align*}
where $R_{p,1}$ is a rotation around the origin that sends $p$ to $p_{1}= (p^0,p^0,0)$, $\gamma_l$ is a parabolic element with fixed point in $1$ that sends the center of $C_l$ to the geodetic joining $0$ to $1$ and finally $\eta_l$ is an hyperbolic element 
with fixed points $\pm 1$ that sends the center of $C_l$ to $0$.
By construction $\Lambda_l$ sends $l$ to a pure energy vector and the remaining kinematical data are sent to some new positions $\Lambda_l k_a$. In the limit $l \to p$ the light ray associated to $\Lambda_l k_a$ moves to the one associated to the 
vector 
\begin{align*}
p_{-1} = \left( \begin{array}{c} 1 \\ -1 \\ 0 \end{array} \right).
\end{align*}
Therefore, what $l$ sees around him is a new light cone with a marked ray corresponding to $p_{-1}$. However the remaining kinematical data $k_a$ may move as $l \to p$, in such a way to compensate the infinite boost $\eta_l$, and in 
this case $l$ would see light rays $k_a$ associated to some ``surviving'' particles.

We can label all these possible limits with exactly the same combinatorial object we needed to label the facets of the Halohedron, arcs on a marked annulus, and thus compactify $\mathcal{K}_{1,n}$ adding factorised components associated to the contraction of the relevant arc. For example to an arc such as the one in the third row of table \ref{tab:facethaloedron} we associate the limit where all particles from $j$ to $i$ fail to remain at a finite position when $l \to p$, and we add to the kinematical space a border of the form $\mathcal{K}_{n_L+1} \times \mathcal{K}_{1,n_R+1}$. 

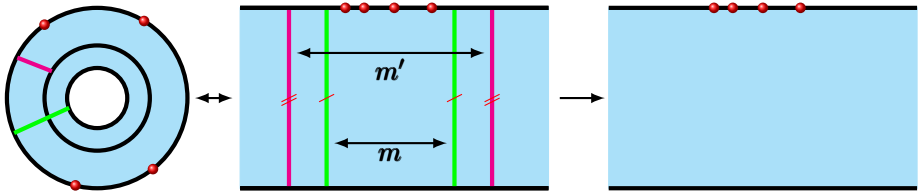
\begin{figure}[!htbp]
\begin{center}
\begin{tikzpicture}[>=latex,scale=0.5, every node/.style={transform shape}]

\begin{scope}
\filldraw[cyan!30!white] (0,0) circle (2.4cm);
\draw [ultra thick] (0,0) circle (2.4cm);
\filldraw[cyan!30!white] (0,0) circle (1.4cm);
\draw [ultra thick] (0,0) circle (1.4cm);
\filldraw[white] (0,0) circle (0.8cm);
\draw [ultra thick] (0,0) circle (0.8cm);
\foreach \x/\y in {-1.4/1.953,1.247/2.05,1.483/-1.891,-0.578/-2.323}
{\shade [ball color=red](\x,\y) circle (4pt);}

\draw[green,ultra thick] (-0.7451, -0.2617) -- (-2.208, -0.9302);
\draw[magenta,ultra thick] (-1.205, 0.7133) -- (-2.124, 1.075);
\end{scope}
\draw [>=latex][thick, <->](2.6,0) -- (3.6,0);
\begin{scope}[xshift=6.2cm]
\filldraw[cyan!30!white] (-2.4,-2.4) rectangle (5.8,2.4);

\draw[ultra thick] (-2.4,2.4) -- (5.8,2.4);
\draw[ultra thick] (-2.4,-2.4) -- (5.8,-2.4);

\draw[green,ultra thick] (-1.2+1.6-0.5,2.4) -- (-1.2+1.6-0.5,-2.4);
\draw[green,ultra thick] (1.2+1.6+0.5,2.4) -- (1.2+1.6+0.5,-2.4);
\draw[red](-1.2+1.6-0.2-0.5,-0.1) -- (-1.2+1.6+0.2-0.5,+0.1);
\draw[red](1.2+1.6-0.2+0.5,-0.1) -- (1.2+1.6+0.2+0.5,+0.1);

\draw[magenta,ultra thick] (-1.2+1.6-1.5,2.4) -- (-1.2+1.6-1.5,-2.4);
\draw[magenta,ultra thick] (1.2+1.6+1.5,2.4) -- (1.2+1.6+1.5,-2.4);
\draw[red](-1.2+1.6-1-0.2-0.5,-0.1) -- (-1.2+1.6-1+0.2-0.5,+0.1);
\draw[red](1.2+1.6+1-0.2+0.5,-0.1) -- (1.2+1.6+1+0.2+0.5,+0.1);
\draw[red](-1.2+1.6-1-0.2-0.5,-0.2) -- (-1.2+1.6-1+0.2-0.5,0);
\draw[red](1.2+1.6+1-0.2+0.5,-0.2) -- (1.2+1.6+1+0.2+0.5,0);

\draw[ultra thick] (-2.4,2.4) -- (5.8,2.4);
\draw[ultra thick] (-2.4,-2.4) -- (5.8,-2.4);

\shade[ball color =red](0.4,2.4) circle (4pt);
\shade[ball color =red](0.9,2.4) circle (4pt);
\shade[ball color =red](1.7,2.4) circle (4pt);
\shade[ball color =red](2.7,2.4) circle (4pt);

\draw[thick, <->](-1.2+1.6+0.2-0.5,-1.2)--(1.2+1.6-0.2+0.5,-1.2) node [midway, below]{\huge $\pmb {m}$};
\draw[thick, <->](-1.2+1.6-1+0.2-0.5,1.2)--(1.2+1.6+1-0.2+0.5,1.2)node [midway, below]{\huge $\pmb {m'}$};

\end{scope}
\draw [>=latex][thick,->] (6.2+5.8+0.3,0) -- (13.5,0); 
\begin{scope}[xshift=16cm]
\filldraw[cyan!30!white] (-2.4,-2.4) rectangle (5.8,2.4);
\draw[ultra thick] (-2.4,2.4) -- (5.8,2.4);
\draw[ultra thick] (-2.4,-2.4) -- (5.8,-2.4);

\shade[ball color =red](0.4,2.4) circle (4pt);
\shade[ball color =red](0.9,2.4) circle (4pt);
\shade[ball color =red](1.7,2.4) circle (4pt);
\shade[ball color =red](2.7,2.4) circle (4pt);

\end{scope}
\end{tikzpicture}
\caption{On the left are shown two annuli with the same punctures but different moduli. They are equivalent to the strips with identifications depicted in the middle, via the map $\sigma \to \exp{\frac{i \sigma}{2 \pi m}}$. The strips both tend to the same surface as the moduli diverge, which is the infinite strip with punctures on the right.}
\label{fig:annuli}
\end{center}
\end{figure}

A comment is in order for the subtle ``cut'' limit that corresponds to an associahedral facet of the kinematical space. This arc represents a situation where all particles survive in the comoving frame of $l$ and therefore we cannot do the limit naively in this way. Accordingly the component that we have to add is not expressed as a factorisation, it is instead $\mathcal{K}_{n+2,\mathrm{forward}}$ which is defined as the set of $n+1$ light rays - the extra light ray is associated to $p$ - up to transformation under the little group of the extra null momenta. We now motivate this: first, suppose $l \to p$ and define $l' := \lambda l$. In a frame comoving with $l$ (and $l'$) the remaining kinematical data $k_a$ tends to fixed position on the boundary of the disk. We have two surfaces $X = \phi(k_a,l)$ and $X' = \phi(k_a,l')$ which can be represented as annuli with the same punctures but different moduli. However, in the limit $l^2 \to 0$ they both degenerate to the same pinched annuli as shown in figure \ref{fig:annuli}.

\begin{figure}[!htbp]
\begin{center}
\begin{tikzpicture}[>=latex,scale=0.75, every node/.style={transform shape}]

\filldraw[cyan!30!white] (0,0) circle (2.4cm);
\filldraw[white] (0,0) circle (0.8cm);
\draw [ultra thick] (0,0) circle (2.4cm);
\draw [ultra thick] (0,0) circle (0.8cm);
\shade [ball color=red] (-2.4,0) circle (4pt);
\shade [ball color=red] (2,1.3) circle (4pt);
\shade [ball color=red] (2,-1.3) circle (4pt);
\foreach \x/\y in {1.595/1.786,-0.6755/2.301,-2.124/1.117,-1.776/-1.627,0.578/-2.323}
{\shade [ball color=red](\x,\y) circle (4pt);}
\draw [red, ultra thick] (2.4,0) -- (0.8,0);
\node at (2.2+0.4,1.5) {$\pmb {i+1}$};
\node at (2.2+0.4,-1.5) {$\pmb i$};
\draw [line width=4pt,<->] (2.4+0.2,0) -- (8-2.4-0.2,0);
\draw [line width=3pt] (2.4+0.2+0.8,-0.4) -- (8-2.4-0.2-0.8,0.4);
\draw [line width=3pt] (2.4+0.2+0.8,0.4) -- (8-2.4-0.2-0.8,-0.4);

\begin{scope}[xshift=8cm]
\filldraw[cyan!30!white] (0,0) circle (2.4cm);
\draw [ultra thick] (0,0) circle (2.4cm);
\draw [ultra thick,red] (0,-2.4) -- (0,2.4);
\shade [ball color=red] (-2.4,0) circle (4pt);
\shade [ball color=red] (2,1.3) circle (4pt);
\shade [ball color=red] (2,-1.3) circle (4pt);
\shade [ball color=red] (-1.386, 1.953) circle (4pt);
\shade [ball color=red] (-1.386, -1.953) circle (4pt);
\node at (1.985+0.4, 1.298){$\pmb +$};
\node at (1.985+0.4, -1.298){$\pmb -$};
\node at (-1.386, 1.953+0.5){$\pmb {i+1}$};
\node at (-1.386, -1.953-0.4){$\pmb i$};
\foreach \x/\y in {-2.11/1.159,-2.333/0.6297,-2.389/-0.3592,-2.221/-0.9441}{\shade [ball color=red](\x,\y) circle (4pt);}
\end{scope}

\end{tikzpicture}
\caption{These facets of and $\mathcal{K}_{1,n}$ and $\mathcal{K}_{n+2}$ are not equivalent.}
\label{fig:wrong}
\end{center}
\end{figure}
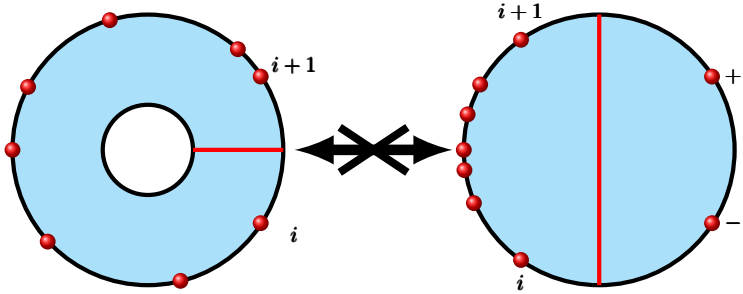

\FloatBarrier
This explains why we consider the light-ray associated with $p$ rather than $p$ itself. Next consider what happens if we take two surfaces with the same $l$ but kinematical data $k' = \Lambda k$ with $\Lambda$ an element of the little group of $l$, which in the disk correspond to an elliptic element with fixed 
point the center of $C_l$. If $l \to p$, and we choose a frame where we see $C_l \to H_p$ while the punctures remain at finite positions,  $\Lambda$ tends to an ideal rotation centered at $p$,  but $X'$ and $X$ are always bi-holomorphic and thus they have the same limit: therefore we have to mod out by parabolic elements of the new light ray.prov
Note that $\mathcal{K}_{n+2,\mathrm{forward}}$ is not the same as the facet of $\mathcal{K}_{n+2}$ labelled by the arc of figure \ref{fig:wrong}, as a simple counting of the dimensions proves.

\FloatBarrier

In conclusion we defined a compactification $\overline{\mathcal{K}}_{1,n}$ of the kinematical space which is combinatorially equivalent to an Halohedron, the map $\phi$ is extended by continuity to a map from the kinematical Haloehdron onto the moduli space Halohedron.

\section{Conclusions and Outlook}
\label{sec:conclusions}

In this paper we started to explore the relations between hyperbolic geometry and positive geometries in moduli spaces of Riemann surfaces.

Using the simplest hyperbolic geometry, the plane, we are directly led to an Associahedron living in the kinematical space and to the tree level scattering equations. We are also guided to think about the momenta of a scattering process as homogeneous 
coordinates on the moduli space $\mathcal{M}_{\mathbf{D},n}$, this fact has an intriguing ``holographical'' flavour, which is in line with other ideas by Pasterski et al \cite{Pasterski:2016qvg}, but ultimately proves to be a problem in extracting the amplitude from the canonical form of the kinematical Associahedron.
This is due to the fact that we cannot find a simple way to gauge fix the invariance under rescaling of the single momenta that allows to make manifest all the boundaries of the kinematical Associahedron.  We already elaborated on the fact that this problem, 
though particularly severe in $1+2$ dimensions, does not seem to be related the peculiarity of the $1+2$ dimensional kinematics, but rather to the fact that we chose a specific dimension: this forces Mandelstam variables to satisfy further relations other than momentum conservation, and the simple gauge fixing of \cite{Arkani-Hamed:2017mur}, which realises the Associahedron as a compact polytope in $\mathbb{R}^N$, fails.

Another dissatisfying fact is the absence in our story of Associahedra related to different cyclic orderings and therefore of the beautiful interpretation of double-partial amplitudes as intersection numbers of these Associahedra, as described in \cite{Mizera:2017cqs,Mizera:2017rqa}.

On the good side, the result of pursuing this approach at loop level is that we land on the Halohedron. The Halohedron vertices and facets can be understood in terms of planar loop Feynman diagrams and cuts of the 1-loop integrand, respectively. The 
fact that tadpoles come in pairs, while the remaining vertices are in 1-1 correspondence with the planar diagrams, is the first promising evidence that the Halohedron is the 1-loop Amplituhedron for the bi-adjoint scalar theory. Moreover the hyperboloid 
model immediately suggests a generalisation of the map $\phi$, that at tree level was equivalent to the scattering equations, to the loop level. It would be interesting to understand whether $\phi$ again gives a solution to the now 1-loop scattering equations, in one of their form \cite{Geyer:2015jch,Cardona:2016gon} and to their higher loops generalisations \cite{Geyer:2016wjx,Geyer:2015bja,Gomez:2016cqb}.

The map $\phi$ has interesting properties. On the one side, this map guides us to identify again an Haloedron in kinematical space, on the other it instructs us to think of the cut condition $l^2 = 0$ in an unusual way: we do not interpret $l^2 = 0$ as ``the loop momentum flowing from particle i to particle i+1 is being cut'', as we usually do when defining loop integrands for color ordered amplitudes, but rather as ``the loop momentum is being cut''.
This is in line with the formula \ref{eq:songformula} proposed by He et al \cite{He:2015yua} for $m^{1-\mathrm{loop}}(1,\dots,n|1,\dots,n)$, where the residue of the integrand at $l^2 = 0$ is manifestly given by a sum over different cyclic ordered amplitudes whose tadpoles cancel in pairs: the sum over the cyclic order is captured by the fact that the locus $l^2 = 0$ is written as the union of different Associahedra and the tadpoles pairing pattern is reproduced by an analogous pairing of the tadpole vertices of the Halohedron.

Admittedly though, the picture is still very speculative and will continue to be until we cannot produce explicit formulae for the amplitude at loop and tree level in the same fashion of \cite{Arkani-Hamed:2017mur}.
This brings us back to the problem encountered at tree level. Since its root seems to lie in the explicit choice of a space time dimension, the most promising direction to solve it is the ``abstract variables'' approach of \cite{Arkani-Hamed:2017mur}. The natural question is then what these variables should parametrise and what their ultimate meaning should be. Whilst far from being able to produce a physically meaningful explanation, hyperbolic geometry seems a good path to find an answer at least on the mathematical side. As 
already mentioned, the hyperbolic geometry approach to the moduli problem goes hand in hand with Teichm\"uller theory, which roughly speaking studies the problem of parametrising the representations of the fundamental group of a topological surface $X$ 
in terms of elements of $\mathrm{Aut}_\mathbf{D}$. In particular, topological identities expressed by $\pi(X)$ must be translated in matrix relations on $\mathrm{Aut}_{\mathbf{D}}$. Consider the simplest case of a sphere with four punctures, then $\pi(X)$ is 
presented as $\langle \gamma_1,\gamma_2,\gamma_3,\gamma_4 | \gamma_1 \gamma_2 \gamma_3 \gamma_4 = 1 \rangle$, $\gamma_i$ being a loop around the $i$-th puncture. The topological identity $\prod_i \gamma_i = 1$ is very similar to a 
momentum conservation relation, and if we define $\gamma_{ij} := \gamma_{i}\gamma_{j}$ it is tempting to make an association
\begin{align*}
\gamma_{ij} \to \mathrm{e}^{k_{ij}},
\end{align*}
so that the usual relations $k_{12} + k_{13} + k_{14} = 0$ is just a re-phrasing of the topological relation of the fundamental group. This analogy suggests that we can try to think of the Mandelstam variables as coordinates on Teichm\"uller space, and 
translates a cubic Feynman diagram in the choice of a chart for this space. This is an interesting line of research that we intend to pursue in future work.


\begin{thebibliography}{19}

\bibitem{ArkaniHamed:2012nw}
  N.~Arkani-Hamed, J.~L.~Bourjaily, F.~Cachazo, A.~B.~Goncharov, A.~Postnikov and J.~Trnka,
  Cambridge University Press. doi:10.1017/CBO9781316091548, 2016.


\bibitem{Arkani-Hamed:2017mur}
  N.~Arkani-Hamed, Y.~Bai, S.~He and G.~Yan,
  arXiv:1711.09102 [hep-th].

\bibitem{Arkani-Hamed:2017tmz}
  N.~Arkani-Hamed, Y.~Bai and T.~Lam,
  JHEP {\bf 1711} (2017) 039
  doi:10.1007/JHEP11(2017)039

\bibitem{Arkani-Hamed:2013jha}
  N.~Arkani-Hamed and J.~Trnka,
  JHEP {\bf 1410} (2014) 030
  doi:10.1007/JHEP10(2014)030


\bibitem{Mizera:2017cqs}
  S.~Mizera,
  JHEP {\bf 1708} (2017) 097
  doi:10.1007/JHEP08(2017)097

\bibitem{Mizera:2017rqa}
  S.~Mizera,
  arXiv:1711.00469 [hep-th].


\bibitem{Cachazo:2013gna}
  F.~Cachazo, S.~He and E.~Y.~Yuan,
  Phys.\ Rev.\ D {\bf 90} (2014) no.6,  065001
  doi:10.1103/PhysRevD.90.065001

\bibitem{Cachazo:2013hca}
  F.~Cachazo, S.~He and E.~Y.~Yuan,
  Phys.\ Rev.\ Lett.\  {\bf 113} (2014) no.17,  171601
  doi:10.1103/PhysRevLett.113.171601

\bibitem{Cachazo:2013iea}
  F.~Cachazo, S.~He and E.~Y.~Yuan,
  JHEP {\bf 1407} (2014) 033
  doi:10.1007/JHEP07(2014)033

\bibitem{Cachazo:2016ror}
  F.~Cachazo, S.~Mizera and G.~Zhang,
  JHEP {\bf 1703} (2017) 151
  doi:10.1007/JHEP03(2017)151


\bibitem{He:2015yua}
  S.~He and E.~Y.~Yuan,
  Phys.\ Rev.\ D {\bf 92} (2015) no.10,  105004
  doi:10.1103/PhysRevD.92.105004


\bibitem{Geyer:2015jch}
  Y.~Geyer, L.~Mason, R.~Monteiro and P.~Tourkine,
  JHEP {\bf 1603} (2016) 114
  doi:10.1007/JHEP03(2016)114

\bibitem{Geyer:2015bja}
  Y.~Geyer, L.~Mason, R.~Monteiro and P.~Tourkine,
  Phys.\ Rev.\ Lett.\  {\bf 115} (2015) no.12,  121603
  doi:10.1103/PhysRevLett.115.121603

\bibitem{Geyer:2016wjx}
  Y.~Geyer, L.~Mason, R.~Monteiro and P.~Tourkine,
  Phys.\ Rev.\ D {\bf 94} (2016) no.12,  125029
  doi:10.1103/PhysRevD.94.125029


\bibitem{Cardona:2016gon}
  C.~Cardona, B.~Feng, H.~Gomez and R.~Huang,
  JHEP {\bf 1609} (2016) 133
  doi:10.1007/JHEP09(2016)133

\bibitem{Cardona:2016bpi}
  C.~Cardona and H.~Gomez,
  JHEP {\bf 1606} (2016) 094
  doi:10.1007/JHEP06(2016)094

\bibitem{Gomez:2016cqb}
  H.~Gomez, S.~Mizera and G.~Zhang,
  JHEP {\bf 1703} (2017) 092
  doi:10.1007/JHEP03(2017)092




\bibitem{delaCruz:2017zqr}
  L.~de la Cruz, A.~Kniss and S.~Weinzierl,
  arXiv:1711.07942 [hep-th].


\bibitem{devadoss} 
  S.~L Devadoss, T. Heath and C. Vipismakul,
  Notices of the American Mathematical Society. 58. (2011). 


\bibitem{abikoff}
  William Abikoff: The Real Analytic Theory of Teichm\"uller Space. AMS, Lecture note in mathematics, Springer-Verlag, 1980



\bibitem{hubbard}
  Teichm\"uller Theory and Applications to Geometry, Topology, and Dynamics: Volume 1: Teichm\"uller Theory, Matrix Editions Ithaca, NY 14850 MatrixEditions.com, 2006


\bibitem{Pasterski:2016qvg}
  S.~Pasterski, S.~H.~Shao and A.~Strominger,
  Phys.\ Rev.\ D {\bf 96} (2017) no.6,  065026
  doi:10.1103/PhysRevD.96.065026


\bibitem{fairlie}
D. B. Fairlie and D. E. Roberts,Dual Models Without Tachyons - A New Approach,unpublished Durham preprint PRINT-72-2440(1972).

\end{thebibliography}

\end{document}